\documentclass[twocolumn,usenatbib]{mnras}
\usepackage{subcaption}
\usepackage{graphicx}
\usepackage{amsmath}
\usepackage{siunitx}
\usepackage{physics}
\usepackage{xfrac}
\usepackage{xspace}
\usepackage{comment}
\usepackage{hyperref}
\usepackage{amssymb}

\def\pmb#1{\setbox0=\hbox{#1}%
    \kern-.025em\copy0\kern-\wd0
    \kern.05em\copy0\kern-\wd0
    \kern-.025em\raise.0433em\box0}

\def\ltsima{$\; \buildrel < \over \sim \;$}
\def\gtsima{$\; \buildrel > \over \sim \;$}
\def\simlt{\lower.5ex\hbox{\ltsima}}
\def\simgt{\lower.5ex\hbox{\gtsima}}
\def\p2Y{\;_2Y}
\def\m2Y{\;_{-2}Y}

\def\mk2{\mu {\rm K}^2}
\def\Planck{\it Planck \rm}

\def\LCDM{$\Lambda{\rm CDM}$ }

\def\LCDMns{$\Lambda{\rm CDM}$}

\def\pmb#1{\setbox0=\hbox{#1}%
     \kern-.025em\copy0\kern-\wd0
     \kern.05em\copy0\kern-\wd0
     \kern-.025em\raise.0433em\box0}

\usepackage{xcolor}

\definecolor{purple}{RGB}{156,81,182}

\begin{document}

\title[Reconstructing the matter power spectrum with cosmic shear]{Reconstructing the matter power spectrum with future cosmic shear surveys}

\author[Preston, Amon \& Efstathiou]{Calvin Preston$^{1,2}$\thanks{E-mail: cp662@cam.ac.uk}, Alexandra Amon $^{3}$\thanks{E-mail: alexandra.amon@princeton.edu}, George Efstathiou $^{1,2}$\thanks{E-mail: gpe@ast.cam.ac.uk} \\
${1}$ Kavli Institute for Cosmology Cambridge,
Madingley Road, Cambridge, CB3 OHA. \\
${2}$ Institute of Astronomy. Madingley Road, Cambridge, CB3 OHA. United Kingdom. \\
${3}$ Department of Astrophysical Sciences, Peyton Hall, Princeton University, Princeton, NJ USA 08544}

\maketitle
\begin{abstract} 
Analyses of cosmic shear typically condense weak lensing information over a range of scales to a single cosmological parameter, $S_8$. This paper presents a method to extract more information from Stage-IV cosmic shear measurements by directly reconstructing the matter power spectrum from linear to non-linear scales. We demonstrate that cosmic shear surveys will be sensitive to the shape of the matter power spectrum on non-linear scales. We show that it should be possible to distinguish between different models of baryonic feedback and we investigate the impact of intrinsic alignments and observational systematics on forecasted constraints. In addition to providing important information on galaxy formation, power spectrum reconstruction should provide a definitive answer to the question of whether weak lensing measurements of $S_8$ on linear scales are consistent with the \Planck \LCDM cosmology. In addition, power spectrum reconstruction may lead to new discoveries on the composition of the dark sector.   
\end{abstract}


\section{Introduction}\label{sec:intro}
Observations of the primary anisotropies and lensing of the cosmic microwave background (CMB) are well described by the \LCDM cosmology, in which the present day energy content is dominated by a positive cosmological constant $\Lambda$ and collisionless cold dark matter \citep{Params:2018,ACT2023}. Measurements of baryon acoustic oscillations \citep[BAO;][]{Alam:2021a, DESI_BAO_2024}, combined with the magnitude-redshift relation of Type 1a supernova \citep{Pantheon_plus} constrain the expansion history and provide strong support for \LCDMns. Combined with CMB data, the parameters of the \LCDM cosmology have been determined to percent-level precision. Overall, it appears that the \LCDM cosmology offers a consistent picture of the background cosmology and perturbations on linear scales.

It is less clear whether the \LCDM cosmology is consistent with observations on small scales where the matter density field is non-linear ($k \simgt 0.1h/\rm{Mpc}$ at the present day), since there are various effects, for example, non-standard dark matter and baryonic physics that can affect structure formation on non-linear scales. If the dark matter is purely collisionless, the non-linear evolution of matter fluctuations can be simulated with N-body simulations to arbitrary accuracy, limited only by the availability of computer power \citep[for a review see][]{Dark_matter_nbody_review}. However, it is now widely accepted that feedback processes associated with galaxy formation can modify the matter power spectrum on small scales \citep[e.g.][]{vanDaalen:2011, Dubois:2014, McCarthy:2017, Springel:2018, Chisari_2019}. Unlike collisionless dark matter, baryonic feedback involves complex physics operating on a wide range of scales and cannot be simulated ab initio. The suppression of the non-linear matter power caused by baryon feedback therefore depends on the specific implementation of `sub-grid' (i.e. unresolved) physics and remains an active area of research \citep[e.g.][]{Schaye2023, Pakamor:2023}. Non-standard dark matter, for example, massive neutrinos, warm dark matter or axions \citep[e.g.][]{Viel, Rogers:2023, Liu:2018, Lague} can produce distinctive features in the matter power spectrum. The matter power spectrum on small scales can therefore provide an important test of the nature of dark matter, provided their signatures can be disentangled from baryonic physics \citep[see, for example,][]{Elbers:2024}.

There is a rich source of information on small scales but the interpretation, via the usual approach of fitting theoretical cosmological models by sampling over parameters, is likely to be difficult since the models themselves are not robust. The purpose of this paper is to present a scheme for reconstructing the matter power spectrum from future cosmic shear surveys in a model independent way. As we will demonstrate, it should become possible to map the power spectrum accurately well into the non-linear regime. If such measurements are found to differ from the non-linear spectrum expected from collisionless dark matter, the shape of the reconstructed spectrum may provide clues to the physics responsible. Furthermore, theorists will be able to explore possible explanations without having to implement the full machinery to perform fits directly to the weak lensing data.

Most analyses of cosmic shear surveys have fitted cosmological parameters to the \LCDM model or simple variants \citep{Asgari:2021,amon:2022,Secco:2022, dalal2023hyper, li2023hyper}. There have been some attempts to constrain the shape of the matter power spectrum from weak lensing data \citep{PKZ_compilation, DouxDESHarmonic, CPAAGPE}. However, because current surveys have limited statistical power, these reconstructions have used coarse bins in wavenumber, are sensitive to assumptions concerning the background cosmology and differ according to whether the reconstructions are designed to recover the linear or non-linear matter power spectra.

Forthcoming surveys, such Vera C. Rubin Observatory’s Legacy Survey of Space and Time\footnote{https://www.lsst.org} (LSST), the ESA’s Euclid mission\footnote{https://www.euclid- ec.org}, and the Roman Space Telescope\footnote{https://roman.gsfc.nasa.gov}, \citep{EuclidForecast, LSSTScience, Roman} are expected to lead to dramatic improvements in the statistical power of cosmic shear measurements. In this paper, we demonstrate that it should be possible to reconstruct the matter power spectrum directly from cosmic shear data over the scale range $10^{-2} < k [h/\rm{Mpc}] < 10^{2}$, i.e. extending well into the non-linear regime. As discussed above, such a reconstruction would provide a powerful probe of baryonic feedback effects and of the nature of dark matter. 

Constraining the matter power spectrum also has implications for the so-called `$S_8$ tension'\footnote{Where $S_8 = \sigma_8 (\Omega_{\rm m}/0.3)^{0.5}$, $\Omega_{\rm m}$ is the present day matter density in units of the critical density, $\sigma_8$ is the root mean square linear amplitude of the matter fluctuation spectrum in spheres of radius $8 h^{-1} {\rm Mpc}$ extrapolated to the present day and $h$ is the value of the Hubble constant $H_0$ in units of $100 \ {\rm km}{\rm s}^{-1}{\rm Mpc}^{-1}$.}. Cosmic shear experiments have consistently reported values of $S_8$ that are up to $\sim 3\sigma$ lower than the value inferred from \Planck for the \LCDM cosmology \citep{Params:2018}. The Kilo-Degree Survey, Dark Energy Survey and the Hyper-Supreme Camera \citep[KiDS, DES, HSC][]{Asgari:2021, amon:2022, dalal2023hyper}, and the combination of DES+KiDS \citep{KiDSDES} all find evidence for low values of $S_{8}$. Various explanations of the $S_8$ tension as a manifestation of a change in the late time physics compared to the early times have been proposed \citep[for a review, see][]{SNOWMASS_tensionsreview} but none have proved compelling. More recently it has been suggested that the $S_8$-tension may be a manifestation of a scale-dependent physics. Specifically, it has been proposed that the \Planck \LCDM model is an accurate description on linear scales, but the matter power spectrum on non-linear scales is suppressed as a consequence of baryonic feedback processes and a possible contribution to the mass density by non-standard dark matter \citep{AAGPE2022, CPAAGPE,Rogers_S8tensionaxions}. These ideas can be tested by reconstructing the matter power spectrum from weak lensing measurements.

This paper is structured as follows. Sec.~\ref{sec:data} describes the generation of fiducial data sets with the characteristics of the LSST Dark Energy Science Collaboration (DESC) Year 1 and Year 10 (Y1/Y10) cosmic shear surveys. Sec.~\ref{sec:methods} describes our method for constraining the suppression of the matter power spectrum on non-linear scales by solving for suppression parameters in discrete bins in $k$-space. Our main results are presented in Sec.~\ref{sec:results}. We compare the accuracy of the reconstructions to simulations of baryonic feedback. We also consider the impact of inaccurate modelling of intrinsic alignment (IA) signals and in removing priors on the background expansion rate. Our conclusions are summarised in Sec.~\ref{sec:conclusion}.

\section{Data}\label{sec:data}

\subsection{DESC-like lensing data vectors}\label{sec:xi_DMO}
\begin{figure}
\centering
\includegraphics[width=1\columnwidth]{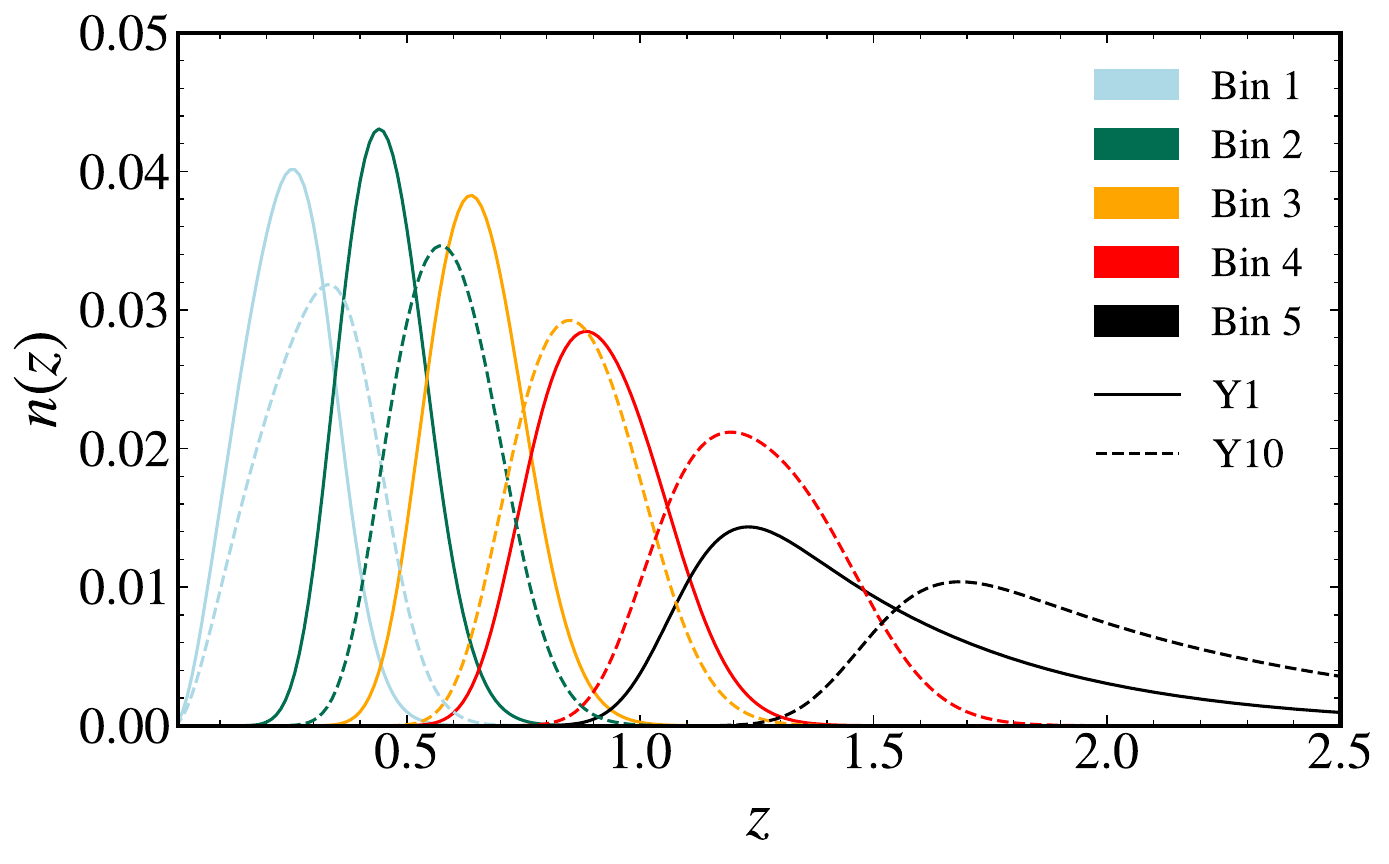} 
\caption{Forecast of the LSST DESC Y1 and Y10 redshift distribution with five redshift bins following \citet{DESC_requirements}. We fix survey parameters for Y1 and Y10 to those given in Table \ref{tab:DESCForecastValues}. Solid lines show forecast Y1 redshift distributions, whilst dashed lines show forecast Y10 distributions. The mean redshift of the Y1 and Y10 samples are approximately 0.85 and 1.05 respectively \citep{DESC_requirements}. For comparison, the mean redshift of DES Y3 is approximately 0.6 \citep{Myles:2021}.}
	\label{fig:LSSTNZ}
\end{figure}

We follow the LSST DESC Science Requirements Document \citep[][hereafter DESC SRD]{DESC_requirements} and assume that shear measurements are divided into five redshift bins that roughly span the redshift range of $0<z<2$ for both Y1 and Y10. The photometric redshift distributions of these bins are set to those given in the DESC SRD and are plotted in Fig.~\ref{fig:LSSTNZ}. Table~\ref{tab:DESCForecastValues} lists the number density of galaxies, $n_{\rm{eff}}$, the shape-noise, $\sigma_{\rm{e}}$\footnote{Where $\sigma_{\rm{e}}$ is the root mean square of the typical intrinsic ellipticity expected from a galaxy in a weak lensing survey \citep{Heymans:2013}} and the survey sky coverage. For comparison, we give the corresponding values for DES Y3 as reported in \citet{amon:2022}.

\begin{table}
\setlength\extrarowheight{5pt}
  \centering
  \caption{Survey statistics forecasted for LSST DESC Y1 and Y10, following \citet{DESC_requirements}. We include DES Y3 for comparison. We report the forecasted number density of galaxies  $n_{\rm{eff}} \, [\rm{galaxies/arcmin}^{2}]$, evenly distributed between the five redshift bins, sky coverage, shape noise $\sigma_{\rm{e}}$ and mean redshift. DES Y3 has no redshift parameters as the calibrated $n(z)$'s of \citet{Myles:2021} are used.}
\begin{tabular}{@{}l|ccccc}
\hline
Parameter & DES Y3 & Year 1 & Year 10 \\
\hline
       \  $n_{\rm{{gal}}}~ (\rm{gal/arcmin^{2})}$ & 5.6 & 10 & 27 &  \\
       \  $\sigma_{\rm{e}}$ & 0.27 & 0.26 & 0.26 & \\
       \  Sky coverage ($\rm{deg}^{2}$) & 5,000 & 18,000 & 18,000 & \\
       \  Mean redshift  & 0.63 & 0.85 & 1.05 & \\
       
\hline
\end{tabular}
\label{tab:DESCForecastValues}
\end{table}
We use these specifications to create simulated DESC Y1/Y10 cosmic shear data vectors consisting of two-point correlation functions, $\xi_{+/-}^{ij}$, for redshift bin pair, $i$ and $j$, assuming the \Planck \LCDM cosmology. Throughout this paper, we use the \textsc{Cosmosis} framework to generate simulated data and to fit parameters \citep{Zuntz:2015}.

The angular convergence power spectra, $C_\kappa^{ij}(\ell)$ as function of multipole $\ell$, are related to the real space correlation functions $\xi_{+/-}^{ij}(\theta)$, via Hankel transforms
\begin{equation}
\xi_{+/-}^{ij} = \frac{1}{2\pi} \int \ell \ J_{0/4}(\ell\theta)\ C_{\kappa}^{ij}(\ell)\ \rm{d} \ell,
\label{equ:totalangspectra}
\end{equation}
where $J_{0/4}$ are the Bessel functions of the first kind. In the small angle (Limber) approximation, the convergence power spectra are related to the matter power spectra, $P_{\rm m}$, according to 
\begin{subequations}
\begin{equation}
C_\kappa^{ij} (\ell) = \int_0^{\chi_{\rm H}} d\chi {q_i(\chi) q_j(\chi) \over [f_K(\chi)]^2 }  P_{\rm m} (\ell/f_K(\chi), \chi), \label{equ:Ckappa1}
\end{equation}
where $\chi$ is the comoving radial distance, $f_K(\chi)$ is the comoving angular diameter distance to radial distance $\chi$ in a Friedmann, Robertson, Walker universe with curvature parameter $K$ and $\chi_H$ is the radial distance to the Hubble scale. Throughout this work, we use the Euclid Emulator \citep{EuclidEmulator} to model the nonlinear matter power spectrum. The lensing efficiency functions $q_i$ are given by
\begin{equation}
    q_i(\chi) = {3H_0^2 \Omega_{\rm m} \over 2c^2} {f_K(\chi) \over a(\chi)} 
    \int_\chi^{\chi_H} d\chi^\prime n_i(\chi^\prime) {f_K (\chi^\prime - \chi) \over f_K(\chi^\prime)},   \label{equ:Ckappa2}
\end{equation}
\end{subequations}
where $a(\chi)$ is the dimensionless scale factor and $n_i(\chi)$ is the redshift distribution in tomographic bin $i$ normalised so that $\int n_i(\chi) d\chi = 1$  \citep[see e.g.][]{Hildebrandt:2017}. In this paper, we assume that the Universe is spatially flat\footnote{Consistent with the strong observational constraint, $\Omega_{\rm{K}} = 0.0004 \pm 0.0018$ \citep{Efstathiou:2020}.} thus $f_K(\chi) = \chi$.

Having specified the redshift distributions and spatial curvature, the correlation functions $\xi_{+/-}$ depend on the coordinate distance $\chi(z)$ as a function of redshift. The parameters $\Omega_{\rm m}$ and $H_0$ affect the amplitude of correlation functions via Eq. \ref{equ:Ckappa2}. If the matter power spectrum $P_{\rm m}(k, z)$ is separable in $k$ and $z$, as expected in linear theory, a deviation in the growth rate from the expectations of \LCDM will suppress the amplitudes of $C_\kappa(\ell)$ and $\xi_{+/-}(\theta)$ without affecting their shapes. Thus, although we assume a \Planck \LCDM cosmology in generating the data vectors and in our default power spectrum reconstruction analysis, this assumption primarily affects the background expansion history and hence $\chi(z)$. Since we solve for the functional form of the power spectrum, the shape and amplitude of the power spectrum is free to deviate from that of the \LCDM cosmology. 

The fact that the non-linear matter power spectrum $P_{\rm m}(k, z)$ is not strictly separable in $k$ and $z$ is discussed in the next section. We will show below that if the cosmological parameters are allowed to vary along with the power spectrum, the constraints on the reconstructed power spectrum are only mildly degraded compared to the case of a fixed \Planck cosmology.

\begin{figure*}
\centering
\includegraphics[width=\textwidth]{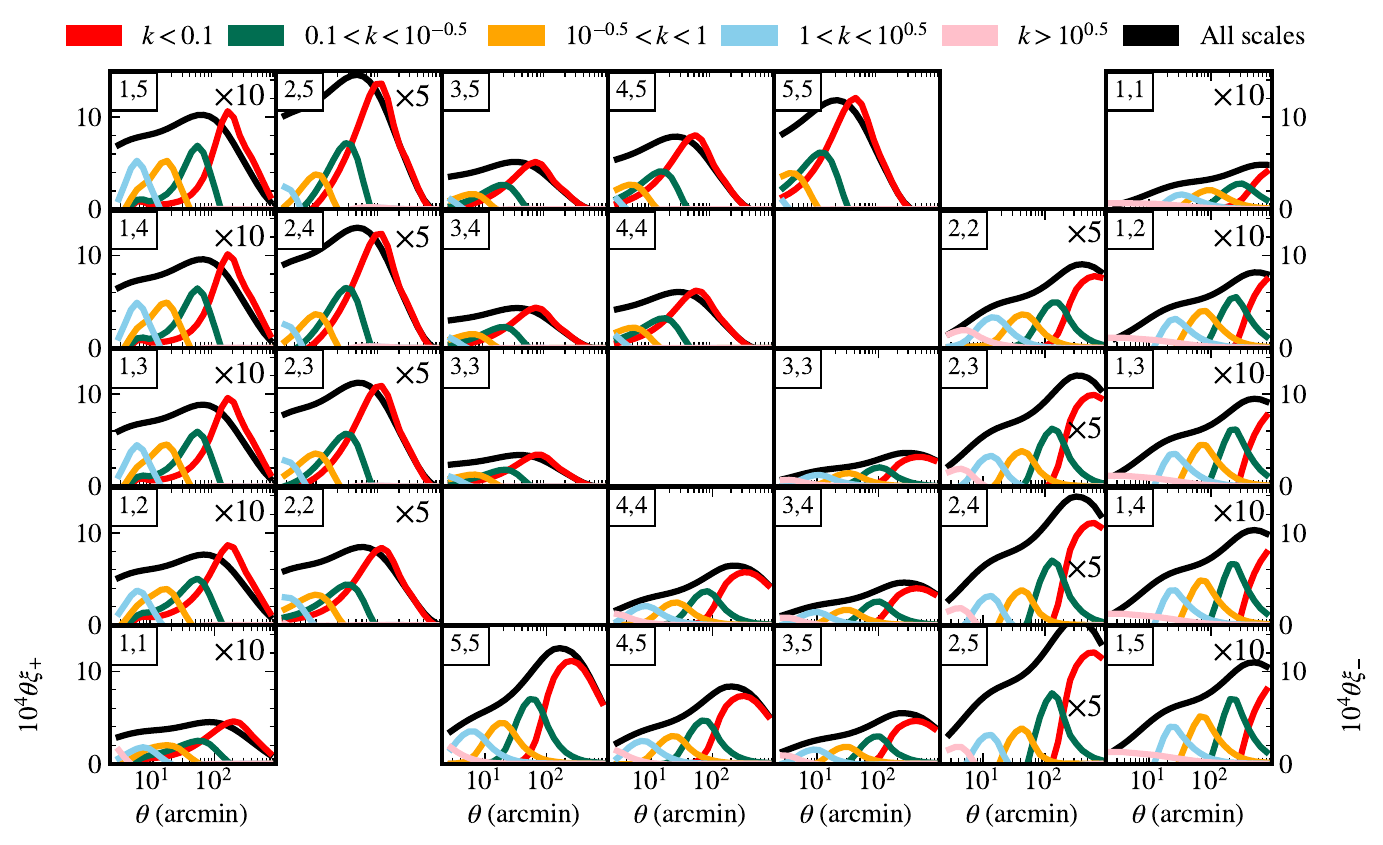} 
    \caption{Forecasted DESC Y10 real space $\xi_{+/-}$ data-vector following \citet{DESC_requirements} fixed at a dark matter only \Planck \LCDM cosmology \citep{Params:2018}. The cross and auto-correlations are shown between each of the five redshift bins. The upper left panels shows $\xi_{+}$, whilst the lower right panels shows $\xi_{-}$, both as a function of angle on the sky, $\theta$. The black lines show the overall data vector, whilst other colours show the contributions from different $k$-scales to the overall fit. Redshift and shear calibration systematics are fixed. Note that scaling factors of $\times 10$ and $\times 5$ have been applied to the $1\times N$ and $2\times N$ correlation functions respectively.}
	\label{fig:LSSTDV}
\end{figure*}

The dark matter only (i.e. no suppression of the non-linear matter power spectrum) correlation functions, $\xi^{ij}_{+/-}$, for the Y10 settings are shown by the black lines in Fig.~\ref{fig:LSSTDV}. We show the contributions from different $k$ scales by lines of different colours. This figure highlights the range of scales that contribute to these measurements that extend will into the non-linear regime. Importantly, we also see a large contribution to the signal from linear scales $k<0.1$, as the cosmic shear signal extends to large $\theta$. 

\subsection{Simulated data specifications}\label{sec:powersuppression}
As discussed in the introduction, baryonic feedback effects are expected to suppress the amplitude of the matter power spectrum on non-linear scales. The predicted form of this suppression is currently uncertain given the sensitivity of hydrodynamical simulations to `sub-grid' physics. One of the main motivations for the work here is to recover the form of the shape of the non-linear power spectrum directly from weak lensing data using as few assumptions as possible. In addition to baryonic feedback, non-standard dark matter could also lead to a suppression of power on small scales. In this paper, we take a phenomenological approach \citep{AAGPE2022} and modify the non-linear matter spectrum by introducing a single parameter $A_{\rm{mod}}$  
\begin{equation}
P_{\rm m}(k, z) =  P^{\rm{L}}_{\rm{m}}(k, z) + A_{\rm{mod}}[P^{\rm{NL}}_{\rm{m}} (k, z) - P^{\rm{L}}_{\rm{m}}(k, z)] \,, \label{equ:Amod}
\end{equation} 
where $P^{\rm{L}}_{\rm{m}}(k, z)$ is the linear theory prediction of the matter power spectrum, $P^{\rm{NL}}_{\rm{m}}(k, z)$ is the dark matter only non-linear prediction and $A_{\rm{mod}}$ controls the amount of power suppression. In our default analysis we choose $A_{\rm{mod}}=0.8$, since this value reconciles weak lensing measurements from DES \citep{amon:2022} with the \Planck \LCDM cosmology \citep{CPAAGPE}. We also consider less extreme suppression, based on the results from current state-of-the-art hydrodynamical simulations \citep[e.g.][]{vandaalen:2020,McCarthy:2017}, to show that it is possible to differentiate between these possibilities using weak lensing measurements.

The intrinsic alignments (IA) of the lensing galaxies contributes to the cosmic shear data vector. The theory of IA is complex and poorly understood \citep[see e.g.][for a review]{Lamman:2024} and so any analysis of IA in this paper will be schematic at best. In Sec.~\ref{sec:IA} we investigate how two widely used models of IA affect reconstructions of the matter power spectrum. We quantify the biases induced if the `wrong' model of IA is used in the reconstruction.

Finally, to illustrate the reconstruction method, we will assume highly idealised data. The redshift distributions have idealised shapes and we ignore any contamination from data calibration issues such as mis-modelled point spread functions. In our analysis, we marginalise over an error on the mean redshift of the distribution for each bin and a shear calibration error for each bin, following the SRD\footnote{The DESC SRD places a Gaussian prior of width $0.003$ on a single shear calibration parameter in each redshift bin and a Gaussian prior in the mean shift redshift of each tomographic bin following $dz_{i}=0.001(1+z_{i})$, where $z_{i}$ is the mean redshift of the $i^{\rm{th}}$ bin. We fix the redshift scatter $\sigma_{z}$ for consistency in comparing to DES Y3.}. Therefore, in total, we include an additional ten nuisance parameters to account for observational systematic uncertainties.

\section{Reconstructing the matter power spectrum}
\label{sec:methods}
Relatively little work has been done on reconstructing the power spectrum from weak lensing data. \citet{PKZ_compilation} and \citet{DouxDESHarmonic} used a variant of the technique described in \citet{Tegmark_Zaldarriaga} to reconstruct the {\it linear} matter power spectrum from DES Y1 and DES Y3 data respectively, assuming the \Planck \LCDM cosmology. In their analyses, the non-linear matter power spectrum is assumed to follow the predictions for collisionless dark matter prediction ignoring suppression caused by baryonic feedback. \citet{CPAAGPE} performed a crude reconstruction of the {\it non-linear} power spectrum from the DES Y3 cosmic shear measurements assuming the \Planck cosmology. This reconstruction demonstrated the extent of the suppression required to explain the $S_8$ tension if the \Planck \LCDM cosmology is correct.

\begin{figure}
\centering
    \includegraphics[width=1\columnwidth]{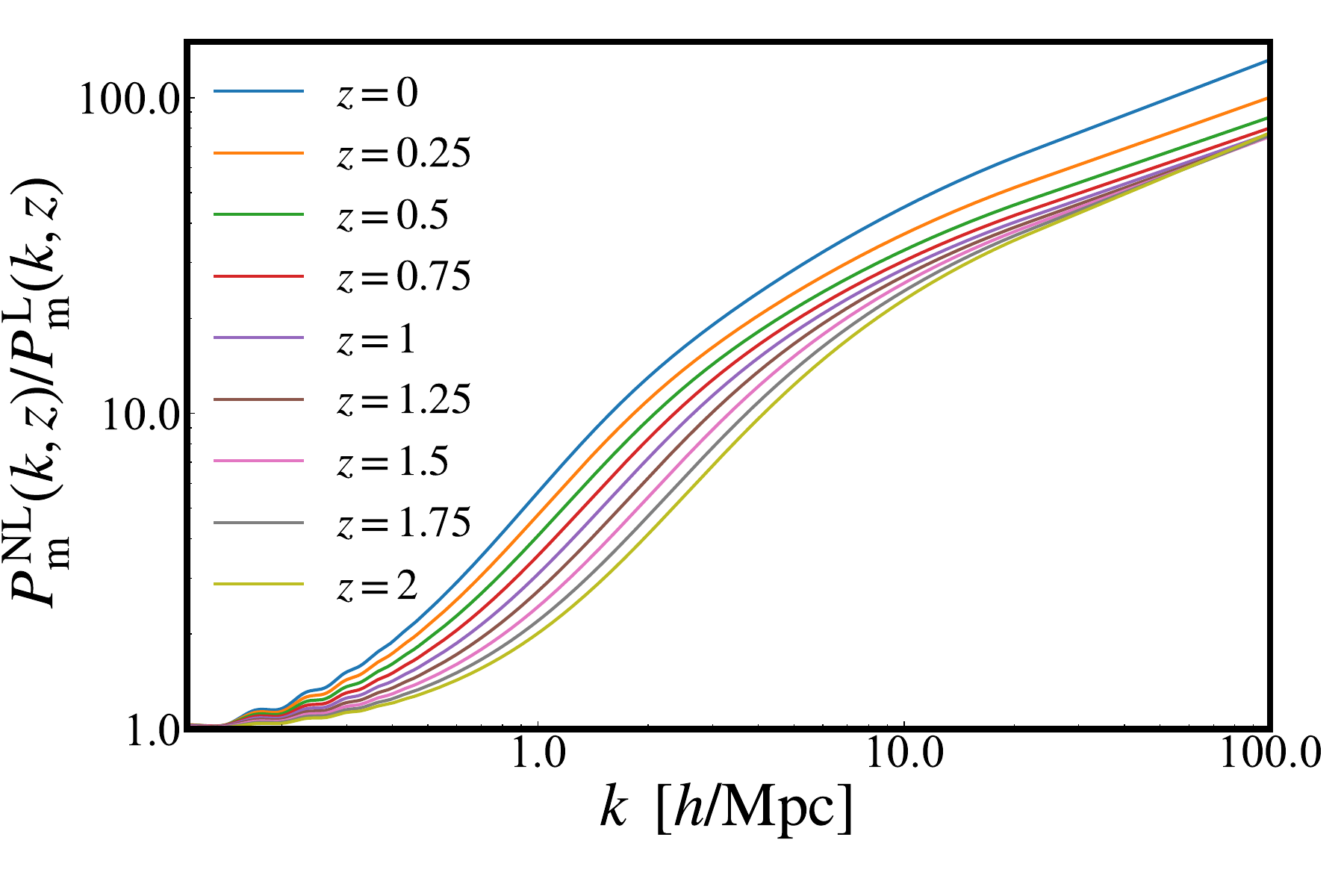} 
\caption{Ratio of the non-linear dark matter only matter power spectrum from Euclid Emulator to the linear matter power spectrum from \texttt{CAMB} \citep{CAMB} from $z=0$ to $z=2$. The increase in amplitude of these curves with decreasing redshift reflect the growth of non-linear structure over cosmic time. In Eq.~\ref{equ:BinningPkz} we make the assumption that linear growth factors do not change over the redshift range we perform our reconstruction.}
\label{fig:NLvsLIN}
\end{figure}

The reconstruction technique presented here is an extension of the method presented in \citet{CPAAGPE}. As noted above, the non-linear matter power spectrum appearing in Eq.~\ref{equ:Ckappa2} is not separable in $k$ and $z$. Fig.~\ref{fig:NLvsLIN} shows the ratio of the non-linear to linear matter power spectrum for a range of redshifts. One way of dealing with the spread shown in Fig.~\ref{fig:NLvsLIN} would be to introduce additional parameters characterising the redshift evolution, but it would be difficult to find a simple parameterisation that could be accurately constrained, even with the statistical power of DESC Y10. Instead we adopt a simpler approach in which we assume the redshift evolution shown in Fig.~\ref{fig:NLvsLIN}, but solve for ratios $\hat{\mathcal{P}}(k_{i})$ in wavenumber bins centred at $k_{i}$:
\begin{equation}
    \frac{P_{\rm{m}}(k,z)}{P^{\rm{DMO}}_{\rm m}(k,z)} = \hat{\mathcal{P}}(k_{i}), 
\label{equ:BinningPkz}
\end{equation}
where $P^{\rm{DMO}}_{m}(k,z)$ is the collisionless dark-matter only non-linear power spectrum. In the limit $P_{\rm{m}}(k,z) = P_{\rm DMO}(k,z)$ the formalism is exact and we would expect to recover $\hat {\mathcal{P}}(k_i) = 1$. In a wide range of cosmological hydrodynamical simulations, the effects of baryonic feedback lead to ratios $\hat{\mathcal{P}}(k_{i})$ that are approximately constant over the redshift range $z \simlt 1$ \citep[see e.g. the compilation of][]{vandaalen:2020}. The weak lensing statistics are sensitive to a relatively narrow range of redshifts, thus in our application to Y10 the suppression parameters $\hat{\mathcal{P}}(k_i)$ quantify the non-linear power spectrum at a redshift $z \sim 0.5$, i.e. half the mean redshift of the Y10 sample.

Following \citet{CPAAGPE}, we choose a set of discrete wavenumber bins $k_i$ and solve for the amplitudes $\hat{\cal P}(k_i)$ by fitting to $\xi_{+/-}$ using the Y1 or Y10 covariance matrices. For the main results in this paper, we use 23 bins, equally spaced in $\log k$ over the range $k=10^{-2}$ to $k=10^{2}$, with open ended-bins at the smallest and largest scales. Similar approaches have been applied to CMB data to reconstruct the primordial primordial curvature spectrum from observations of the CMB temperature and polarisation anisotropies \citep[e.g.][]{Bridle:2003, Peiris:2010, Planck2018Inflation, Handley:2019}. In the case of the CMB, the power spectrum reconstructions can become unstable if the binning in k-space is chosen to be too fine. The solutions can be stabilised by penalising reconstructions with large gradients. In our applications, we found stable results using 23 bins without the need to apply gradient penalties. Since we are searching for smooth variations of the power spectrum, there is no compelling reason to use finer bins. Appendix~\ref{sec:binning} discusses the application of gradient penalties and demonstrates that our results are insensitive to the choice of bin widths.

In our baseline analysis, we introduce 23 free parameters $\hat{\cal P}(k_i)$ as described above and fix the background geometry to the \Planck \LCDM described in Sec.~\ref{sec:data}. We apply the  \texttt{MultiNest} algorithm \citep{Feroz_2009} to sample the parameter space, fitting $\xi_{+/-}$ for a chosen value of the suppression parameter $A_{\rm mod}$ of Eq.~\ref{equ:Amod}. As stressed above, since we are solving for the shape of the power spectrum, the reconstructed spectrum is free to differ from the shape expected for the \Planck \LCDM cosmology. The assumption of a spatially flat \Planck \LCDM cosmology fixes the geometry and background expansion history and therefore the relation between the radial distance $\chi$ and redshift. In our formulation of Eq.~\ref{equ:BinningPkz}, the growth rate of the power spectrum is assumed to follow that of collisionless dark matter in the \LCDM cosmology. Section~\ref{sec:free_cosmology} shows how the power spectrum reconstruction is degraded if the cosmological parameters are determined from the lensing data alone, i.e. with no use of external data. These specific forecasts are likely to be pessimistic, since external data from surveys such as DESI \citep{DESI_BAO_2024} in combination with Type IA supernovae \citep{Brout:2021} will strongly constrain the expansion history.
 
\section{Results}
\label{sec:results}

\begin{figure}
\centering
    \includegraphics[width=1\columnwidth]{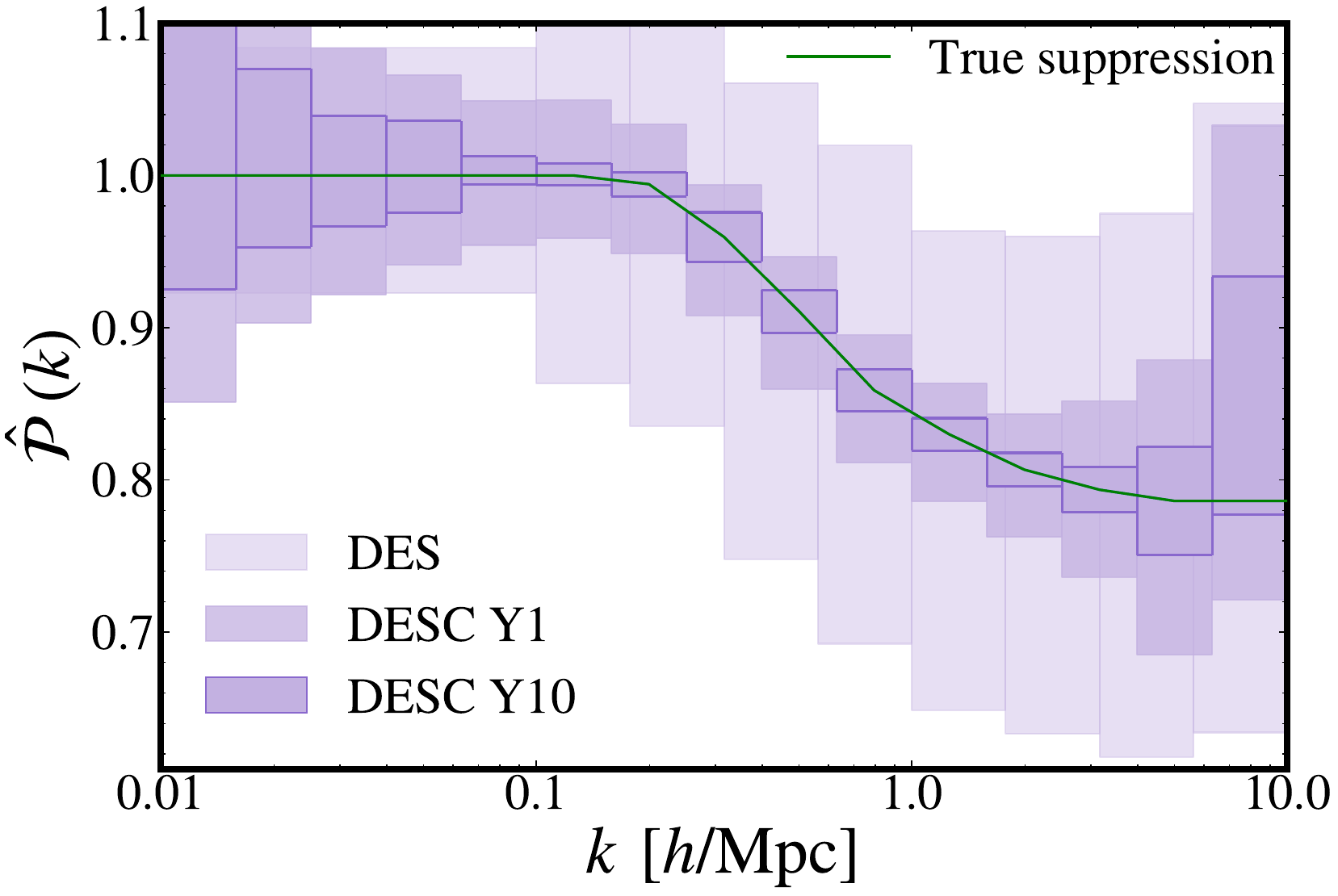} 
\caption{Forecast of LSST DESC Y1 and Y10 constraints on the suppression of the non-linear matter power spectrum relative to the dark matter-only spectrum predicted at $z=0$. The darkest, boxed regions show the Y10 constraints, whilst the next darkest region shows constraints for Y1. For comparison, the lightest shaded regions shows the constraining power of DES. For DESC Y1 and Y10 we select 23 equally spaced bins in $\log (k)$ between $10^{-2}<k<10^{2}$ (except for the open-ended bins at the smallest and largest scales), whilst for the less constraining DES Y3 data we choose 10 bins in $\log (k)$ between $10^{-1}<k<10$. The suppression assumed in creating the data vector is given by the green line.}
\label{fig:PKSAMPLEY1vsY10vsDES_SUPPRESSION}
\end{figure}

In our default analysis, we assume the non-linear power spectrum model of Eq.~\ref{equ:Amod} with a suppression parameter $A_{\rm mod}=0.8$. A non-linear suppression of this order reconciles the \Planck\ \LCDM cosmology with the KiDS and DES Y3 weak lensing measurements \citep{CPAAGPE} and serves as a useful target for our analysis. The results of applying our reconstruction method to DESC Y1 and DESC Y10 simulations are shown in Fig.~\ref{fig:PKSAMPLEY1vsY10vsDES_SUPPRESSION}. For comparison, we show simulated reconstructions for the fiducial DES Y3 measurements computed in the same way as the DESC simulations, but using the redshift distributions and shape noise noise levels from \cite{Myles:2021} and \cite{amon:2022} (see Table~\ref{tab:DESCForecastValues}). Since the constraining power of DES Y3 cosmic shear is significantly weaker than DESC, we use 10 bins in wavenumber evenly spaced between $k = 10^{-1} h/\rm{Mpc}$ and $k = 10 h/\rm{Mpc}$, with open-ended bins at the largest and smallest scales. The DES Y3 results agree well with the power spectrum reconstruction using the real DES Y3 data shown in Fig. 6 of \cite{CPAAGPE}. In contrast to DES, the DESC forecasts have much higher signal-to-noise. Even with DESC Y1, the input power spectrum is recovered to high accuracy demonstrating that a non-linear suppression with $A_{\rm{mod}}\approx 0.8$ is easily detectable.

The most accurately constrained scales are the mildly non-linear scales of $10^{-1} h/{\rm{Mpc}} < k <$ 10 $h/\rm{Mpc}$. At high wavenumbers, the constraints are limited by minimum angular scale measured, $\theta=2.5$ arcmin, following \citet{DES:2021}. The errors on the linear scales are set by the size of the surveys. The constraints on intermediate scales are especially important, however, since we expect to see the effects of baryonic feedback on these scales. Hydrodynamical simulations are most discrepant with each other over this range of wavenumbers \citep[see e.g.][]{vandaalen:2020}.

\subsection{Impact of observational systematics}
\begin{figure}
    \centering
    \includegraphics[width=1\columnwidth]{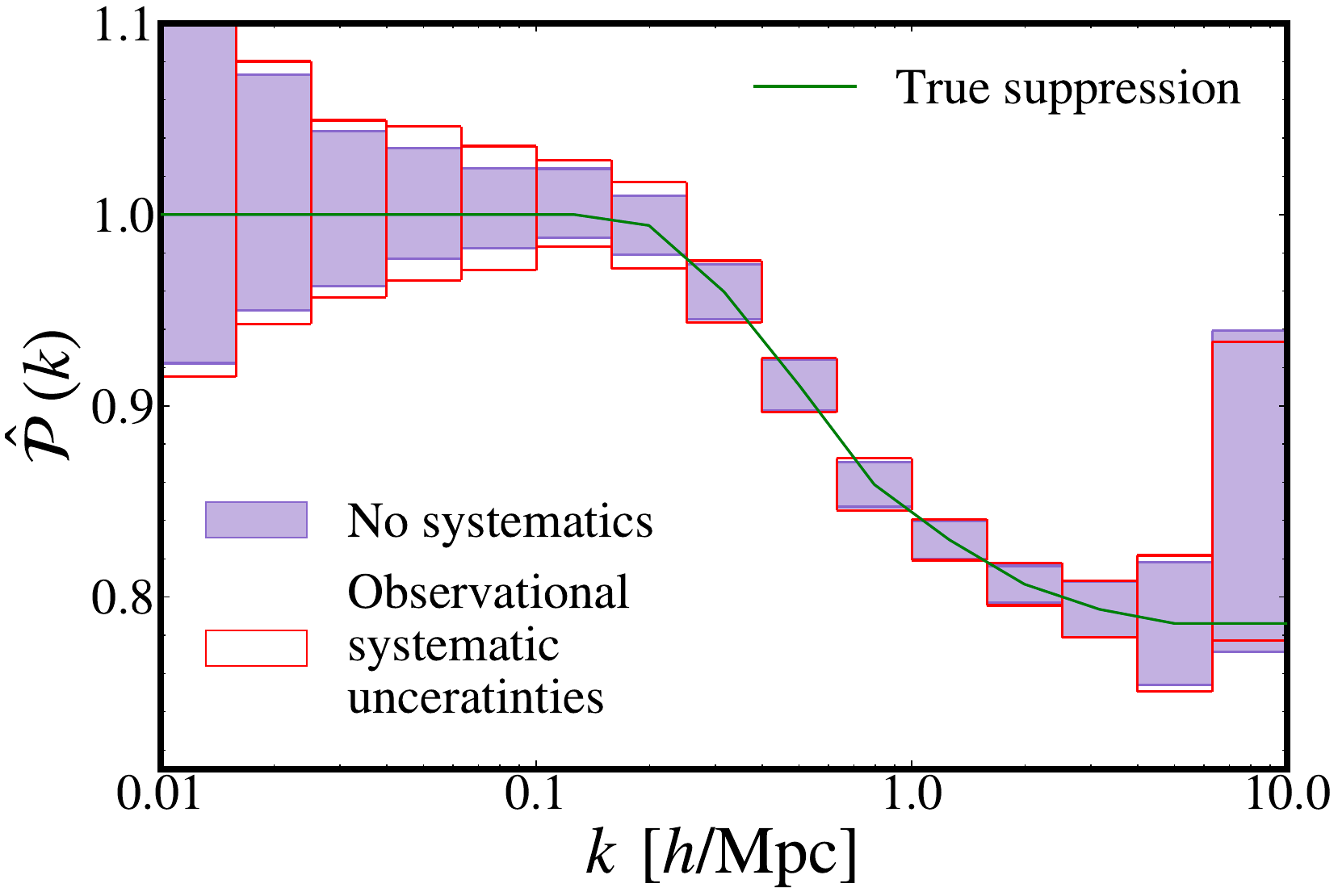} 
\caption{Impact of observational systematics on the power spectrum reconstruction forecast for DESC Y10. The open boxes include nuisance parameters to account for the systematic uncertainties on the mean redshift of each bin and the shear calibration, as prescribed in the DESC SRD, while the filled purple boxes show the result without them, as in Fig.~\ref{fig:PKSAMPLEY1vsY10vsDES_SUPPRESSION}. }
\label{fig:systematics_comparison}
\end{figure}

The SRD for DESC Y1 and Y10 \citep{DESC_requirements} places threshold requirements for the precision of mean shear-shape uncertainty and redshift calibration parameters in each tomographic bin of the cosmic shear analysis. We include these as priors in our analysis. The results of fixing these calibration parameters compared to following the requirements of the SRD are shown in Fig.~\ref{fig:systematics_comparison}. We see that the uncertainty in our results are not systematics dominated.

\subsection{Testing models of baryonic feedback}
\begin{figure*}
    \centering
    \begin{minipage}{.49\textwidth}
    	\includegraphics[width=\columnwidth]{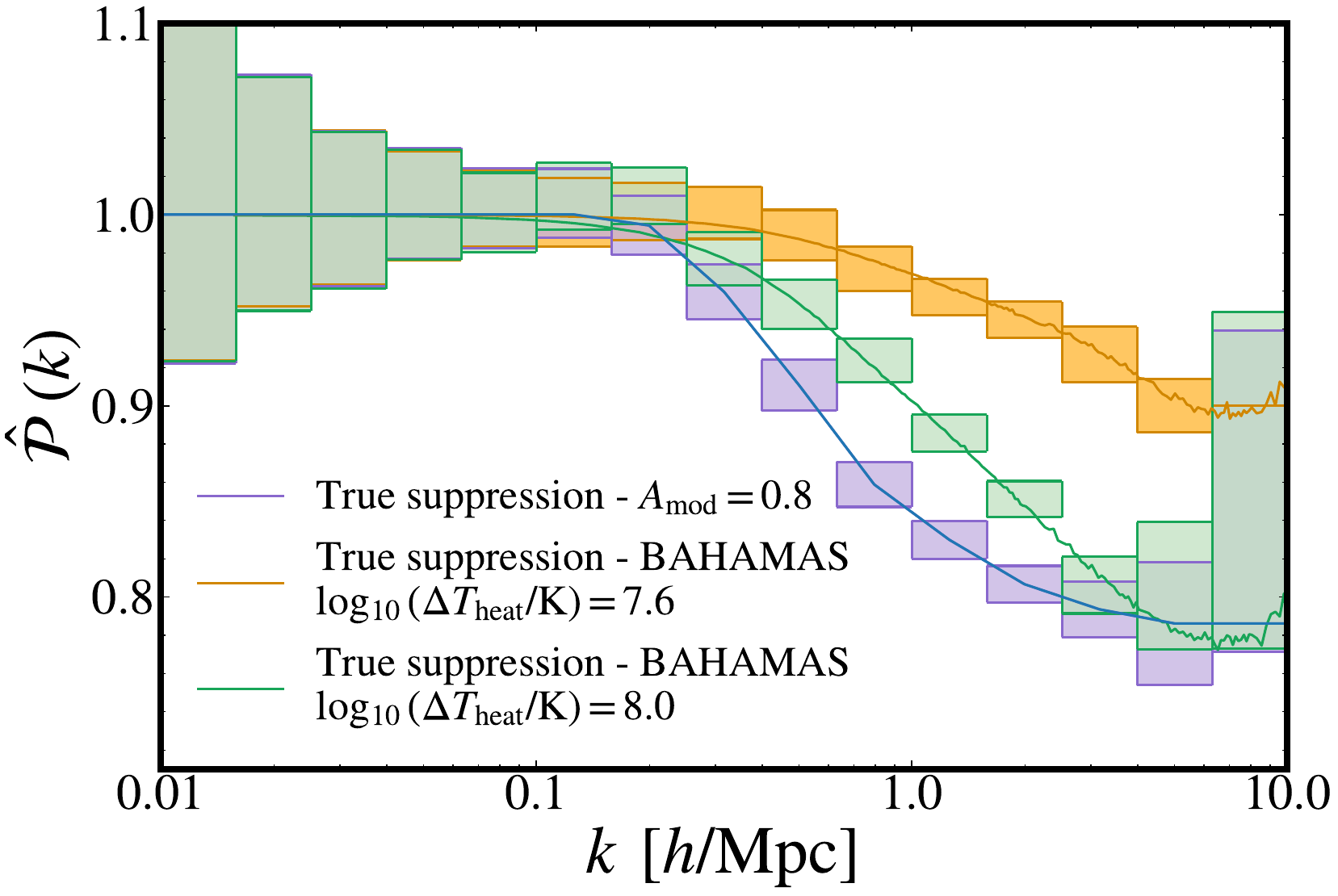} 
    \end{minipage}
    \begin{minipage}{.49\textwidth}
    \centering
    	\includegraphics[width=\columnwidth]{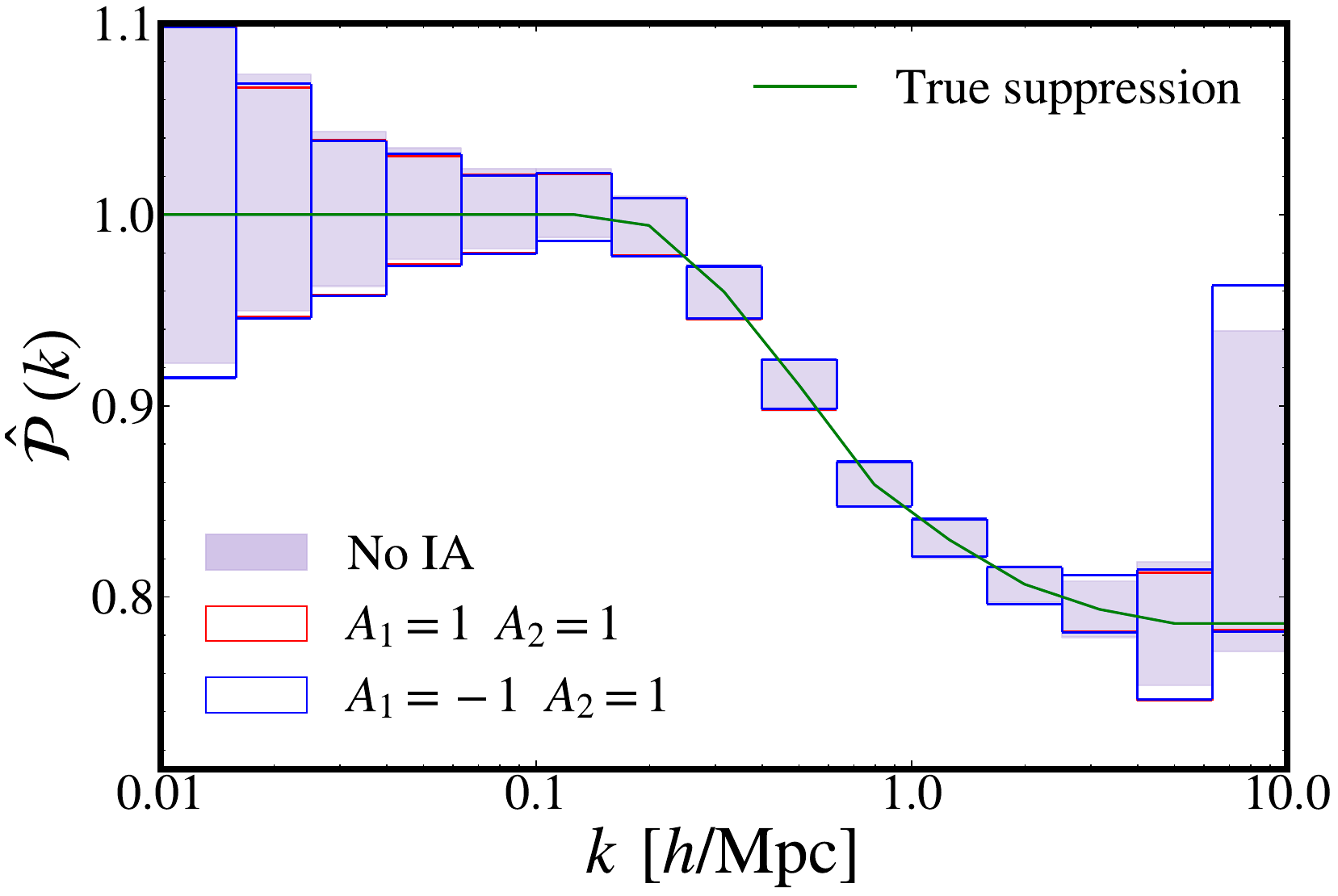} 
    \end{minipage}
\caption{Left: The power of DESC Y10 in recovering different models of power suppression. Here we give three examples: the $A_{\rm{mod}}$ model of \citet{AAGPE2022} and the baryonic feedback prescription of the BAHAMAS hydrodynamical simulations \citep{BAHAMAS}, for two different $\log_{10}( \Delta T_{\rm{heat}}/\rm{K})$ parameters. We see that their is significant enough constraining power to disentangle the three different prescriptions in the approximate scale range $0.2 < k < 2$. Right: Simulated data at DESC Y10 settings following Table \ref{tab:DESCForecastValues} with IA's created with the different amplitudes of the TATT model (with other redshift and bias parameters sets to 0). The transparent boxes are results of analysing this TATT data with the more simplistic NLA model, a subspace of the TATT model. In purple are the year constraints from DESC Y10 cosmology with no IA (as shown in Fig.~\ref{fig:PKSAMPLEY1vsY10vsDES_SUPPRESSION}). We see that, whilst the largest and smallest scales are biased by modelling with an incorrect IA system, they are still generally consistent with the true suppression results. The most tightly constrained scales, $10^{-1} < k < 10$, are well recovered by our method. These results are made with fixed observational systematics (redshift and shear calibration).}
\label{fig:Baryons_IA_Pk}
\end{figure*}

Constraints from cosmic shear surveys offer an extremely promising way of constraining baryonic physics \citep[see, for example][]{Bigwood:2024}. The advantage of the approach proposed here is to constrain the power suppression in a model-independent way. This is illustrated in the left hand panel of Fig.~\ref{fig:Baryons_IA_Pk} showing power spectrum reconstructions for DESC Y10. The purple squares show the $A_{\rm mod}=0.8$ reconstruction as plotted in Fig.~\ref{fig:PKSAMPLEY1vsY10vsDES_SUPPRESSION}. The orange and green squares show reconstructions from data vectors constructed with power suppression following the BAHAMAS model \citep{BAHAMAS} for two values of their feedback parameter $\log_{10}( \Delta T_{\rm{heat}}/\rm{K})= 7.6$ (their favoured value) and $\log_{10}( \Delta T_{\rm{heat}}/\rm{K})= 8.0$. These models are easily distinguishable by DESC Y10. At wavenumbers $k \simgt 7 h/{\rm Mpc}$ the two BAHAMAS models show an upturn in the spectrum. This `spoon-like' shape (which is not described by the phenomenological $A_{\rm mod}$ model) is characteristic of hydrodynamical simulations and is caused by the cooling of baryons and star formation associated with galaxy formation on small scales. The predicted upturn could, in principle, be seen if reliable cosmic shear measurements can be made on scales $\simlt 2.5 \ {\rm arcmin}$.

\subsection{Impact of modelling intrinsic alignments} \label{sec:IA}

So far, we have assumed that galaxies are oriented at random and so the lensing power spectrum is given by Eq.~\ref{equ:Ckappa1}. Intrinsic alignments (IA) could lead to additional contributions to the cosmic shear power spectrum complicating the interpretation of weak lensing measurements \citep[see e.g.][and references therein]{Joachimi:2015, Lamman:2024}. To extract an unbiased reconstruction of the matter power spectrum, it is necessary to model and correct for IA. The total IA signal is comprised of two distinct components:
\begin{enumerate}
    \item Intrinsic-Intrinsic galaxy alignments (II correlations) are caused by correlations in the orientations of galaxies generated during the galaxy formation processes \citep{Crittenden}.
    \item Galaxy-Intrinsic alignments (GI correlations) arise through the cross-correlation of the intrinsic ellipticity of a galaxy and its cosmic shear signal \citep{Hirata&Seljak} as a consequence of correlations between galaxy shapes and the surrounding density field. In some models of IA, the GI effect dominates over the II correlations leading to a reduction of the measured shear-shear signal. 
\end{enumerate}
II signals have been detected at high significance in several surveys of early type galaxies \citep[e.g.][]{BrownIA,KiDSIA}. The GI term has also been detected by \cite{Mandelbaum2006} from a sub-sample of the Sloan digital Sky Survey \citep[SDSS][]{York:2000}. Both the II and GI contributions are thought to depend on galaxy type, (predicted to be stronger for early type galaxies), and are therefore sensitive to the selection criteria of the cosmic shear sample.

Including intrinsic alignments in the total observed angular power spectrum, we can write the power spectrum appearing in Eq.~\ref{equ:totalangspectra}) as the sum of three terms:
\begin{equation}
C^{ij}_{\rm{Total}}(l) = C^{ij}_{\rm{GG}}(l) + C^{ij}_{\rm{GI}}(l) + C^{ij}_{\rm{II}}(l), 
\label{equ:angpowerspectratotal}
\end{equation}
where the pure cosmic shear signal $C^{ij}_{\rm GG}$, is related to the matter power spectrum according to Eq.~\ref{equ:Ckappa1}. There is, however, no exact theory to relate the II and the GI terms to the matter power spectrum, and so we will follow the weak lensing literature and write, assuming a spatially flat geometry, 
\begin{subequations}
\begin{equation}
    C^{ij}_{\rm{II}}(l) = \int_{0}^{\infty} \frac{n^{i}(\chi)n^{j}(\chi)}{\chi^{2}} P_{\rm{II}}(k,z) \rm{d}\chi,
\label{equ:limberapproximation_II}
\end{equation}
\begin{equation}
    C^{ij}_{\rm{GI}}(l) = \int_{0}^{\infty} \frac{[g^{i}(\chi)n^{j}(\chi)+n^{i}(\chi)g^{j}(\chi)]}{\chi^{2}} P_{\rm{GI}}(k,z) \rm{d}\chi,
\label{equ:limberapproximation_GI}
\end{equation}
\end{subequations}
and apply a simplified model to relate the power spectra $P_{\rm{II}}(k,z)$ and $P_{\rm{GI}}(k,z)$ to the matter power spectrum.

The most widely used model is the non-linear alignment model \citep[NLA,][]{Bridle_KingNLA}. The model is based on the ansatz that intrinsic shear of a galaxy is proportional to the linear tidal field \citep{Catelan:2001,Hirata&Seljak}. To extend the model into the non-linear regime, \cite{Bridle_KingNLA} replaced the linear matter power spectrum with the non-linear matter power spectrum, leading to the expressions:
\begin{subequations}
\begin{equation}
P_{\rm{II}}(k,z) = A(z)^{2}P_{\rm{m}}(k,z), 
\label{equ:NLA_II}
\end{equation}
\begin{equation}
P_{\rm{GI}}(k,z) = A(z)P_{\rm{m}}(k,z), 
\label{equ:NLA_GI}
\end{equation}
\end{subequations}
where redshift dependent function $A(z)$ is given by
\begin{equation}
A(z)=-A_{1}C_{1}\rho_{\rm{crit}}\frac{\Omega_{\rm{m}}}{D(z)}\left( \frac{1+z}{1+z_{0}} \right)^{\eta_{1}}, 
\label{equ:NLA_A}
\end{equation}
$\rho_{\rm{crit}}$ is the critical density at the present day and $D(z)$ is the growth rate of linear fluctuations normalised to unity at the present day. The constant $C_1$ is conventionally set to $C_1 = 5 \times 10^{-14}$${M_{\odot}}$Mpc$^2$$/h^2$  to match the II measurements of \citet{BrownIA}. This version of the NLA model has two free parameters, the amplitude $A_{1}$ and the parameter $\eta_{1}$ which describes the redshift evolution relative to a pivot redshift $z_{0}$ (here fixed to $z_{0}=0.62$ to match the value chosen for the DES Y1 shear analysis \citep{Samuroff2019}). Some variants of the NLA model include a dependence on the luminosity of source sample \citep[e.g.][]{Joudaki:2017cfht}.

The 5-parameter tidal alignment and tidal torquing model \citep[TATT,][]{BlazekTATT} extends above model to include terms quadratic in the tidal field. The intrinsic galaxy shear is expanded as
\begin{equation}
\gamma_{ij}^{\rm{I}} = B_{1}s_{ij} + B_{2} \left(s_{ik}s_{kj} - \frac{1}{3}\delta_{ij}s^{2} \right) + B_{1{\delta}}(s_{ij} \delta), 
\label{equ:IntrinsicGalaxyShape}
\end{equation}
where $s_{ij}$ is the linear tidal field and $\delta$ is the density field. The linear term, proportional to $B_{1}$, describes tidal alignments. Consistency with the NLA model requires $B_{1}(z) = A(z)$, where $A(z)$ is given by Eq.~\ref{equ:NLA_A}. The $B_{2}$ term describes tidal torques and is given by
\begin{equation}
B_{2}(z)=5A_{2}C_{1}\rho_{\rm{crit}}\frac{\Omega_{\rm{m}}}{D(z)}\left( \frac{1+z}{1+z_{0}} \right)^{\eta_{2}},
\label{equ:TATT_B2}
\end{equation}
\citep{BlazekTATT}. The term proportional to $B_{1{\delta}}$ accounts for the fact that galaxy shear surveys sample shear at biased positions in the density field. If galaxies are linearly biased with respect to the density field, then $B_{1 \delta}=B_{1}\times b$ where $b$ is the galaxy bias. In its simplest form, the TATT model has five free parameters $A_{1}, A_{2}, \eta_{1}, \eta_{2}$ and $b$. The expressions for the TATT contributions to the total lensing power spectrum are complicated and can be found in Sections A-D of \cite{BlazekTATT}. In this paper, we use the implementation of TATT in \textsc{Cosmosis}.

\begin{figure}
    \centering
    \includegraphics[width=1\columnwidth]{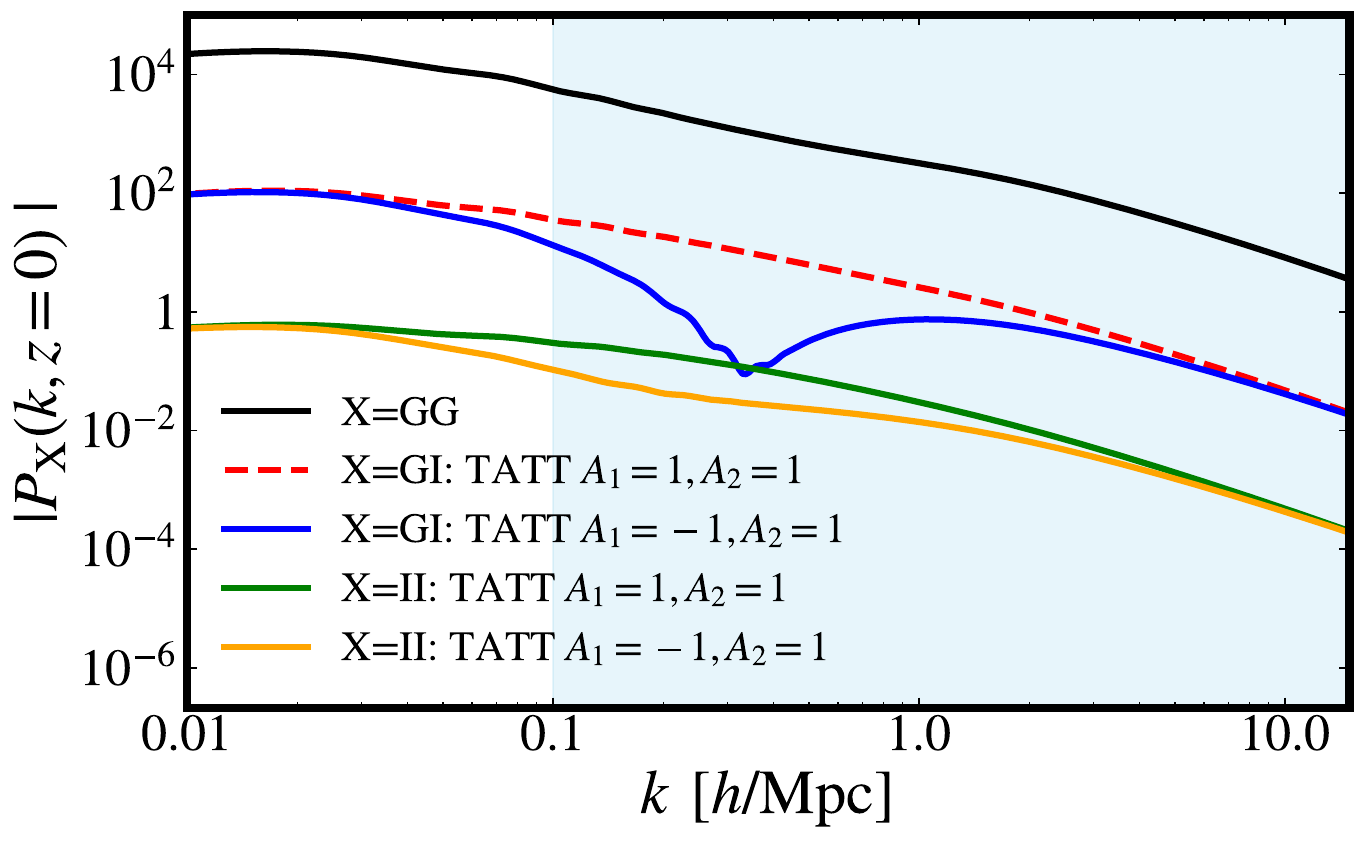} 
\caption{Contributions to power spectra for different values of the IA parameters, within the context of the TATT model. The matter power spectrum, GG, is plotted in black. The GI components, which dominate over II, for different TATT amplitudes are plotted in red and blue. The GI contribution for $A_{1}=1$ and $A_{2}=1$ is negative and we plot the modulus as the dashed line. The shaded blue region indicates the approximate scales that may be affected by baryonic feedback.}
\label{fig:tatt_gi_ii}
\end{figure}

As evident from Fig.~\ref{fig:LSSTDV}, most of the statistical power of weak lensing surveys comes from scales that are well into the non-linear regime. Neither the NLA or the TATT have a sound theoretical basis in this regime. Attempts have been made to model IA including higher order terms using either effective field theory of non-linear structure formation \citep[e.g.][]{Bakx2023}, semi-numerical methods \citep[e.g.][]{Maion_IA_Perturbative}, or the halo model \citep[e.g.][]{Fortuna2020} but these have not yet been used in the analysis of a major cosmic shear survey. Typically, analyses have adopted the NLA model \citep[e.g.][]{Asgari:2020} or performed their analysis with both NLA and TATT \citep{Secco:2022,amon:2022,dalal2023hyper, KiDSDES}, finding shifts in the value of $S_8$ of $\sim 0.5\sigma$.

To illustrate the effects of IA, Fig.~\ref{fig:tatt_gi_ii} shows the spectra $P_{\rm{m}}(k, z)$, $P_{\rm{II}}(k, z)$ and $P_{\rm{GI}}(k, z)$ appearing in Eqs.~\ref{equ:Ckappa1}, \ref{equ:limberapproximation_II} and \ref{equ:limberapproximation_GI} in the TATT model for several values of the parameters $A_1$ and $A_2$ with $\eta_1$ and $\eta_2$ set to zero and ignoring the $B_{1\delta}$ term in Eq.~\ref{equ:IntrinsicGalaxyShape}. For reference, the peak posterior values found from DES Y3 cosmic shear with `optimized' scale-cuts are $A_1 \approx -0.5$, $A_2 \approx 1$ (see Fig. 15 of \cite{amon:2022}, though the posteriors are so broad that they cannot exclude values of zero for both $A_1$ and $A_2$. As discussed by \cite{Hirata&Seljak}, the GI term can dominate over the II term, though for the parameters shown in Fig.~\ref{fig:tatt_gi_ii}, both terms are small in comparison to the GG term. Nevertheless, in the era of precision cosmology, inaccurate modelling of IA can lead to biases in cosmological parameters \citep{TroxelIA}. Given the lack of understanding of IA, it is not possible to make a comprehensive assessment of the impact of IA on power spectrum reconstructions. Instead, we assess the biases caused by using the NLA model to analyse data vectors generated with the more complex TATT model for the values of $A_1$ and $A_2$ shown in Fig.~\ref{fig:tatt_gi_ii}.

Results of this test are shown in the right hand plot of Fig.~\ref{fig:Baryons_IA_Pk} for the fiducial case of $A_{\rm{mod}}=0.8$. Evidently, for these IA parameters, the NLA model has sufficient flexibility to absorb the IA contributions of the TATT model leading to negligible bias in the power spectrum reconstruction. Given the relative contributions of the GG, GI and II power spectra shown in Fig.~\ref{fig:tatt_gi_ii}, it seems unlikely that IA's could seriously bias power spectrum reconstructions at wavenumbers $k \simgt  5 \ h/{\rm Mpc}$, where numerical simulations suggest power suppression of more than about 10\% caused by baryonic feedback. However, it would be premature to conclude on the basis of our tests that IA modelling does not bias reconstructions on scales $k \simlt  0.3 \ h/{\rm Mpc}$ where the power spectrum suppression from baryon feedback is expected to be at the level of a few percent. 

Looking to the future, further work is needed to understand model the impact of IA, guided by direct measurements from spectroscopic surveys and numerical simulations \citep[e.g.][]{Tenneti:2015,Hilbert:2017, Johnston_2019, Samuroff_2023}.

\subsection{Constraining cosmology and the non-linear matter power spectrum} \label{sec:free_cosmology}

\begin{figure}
\centering
    \includegraphics[width=1\columnwidth]{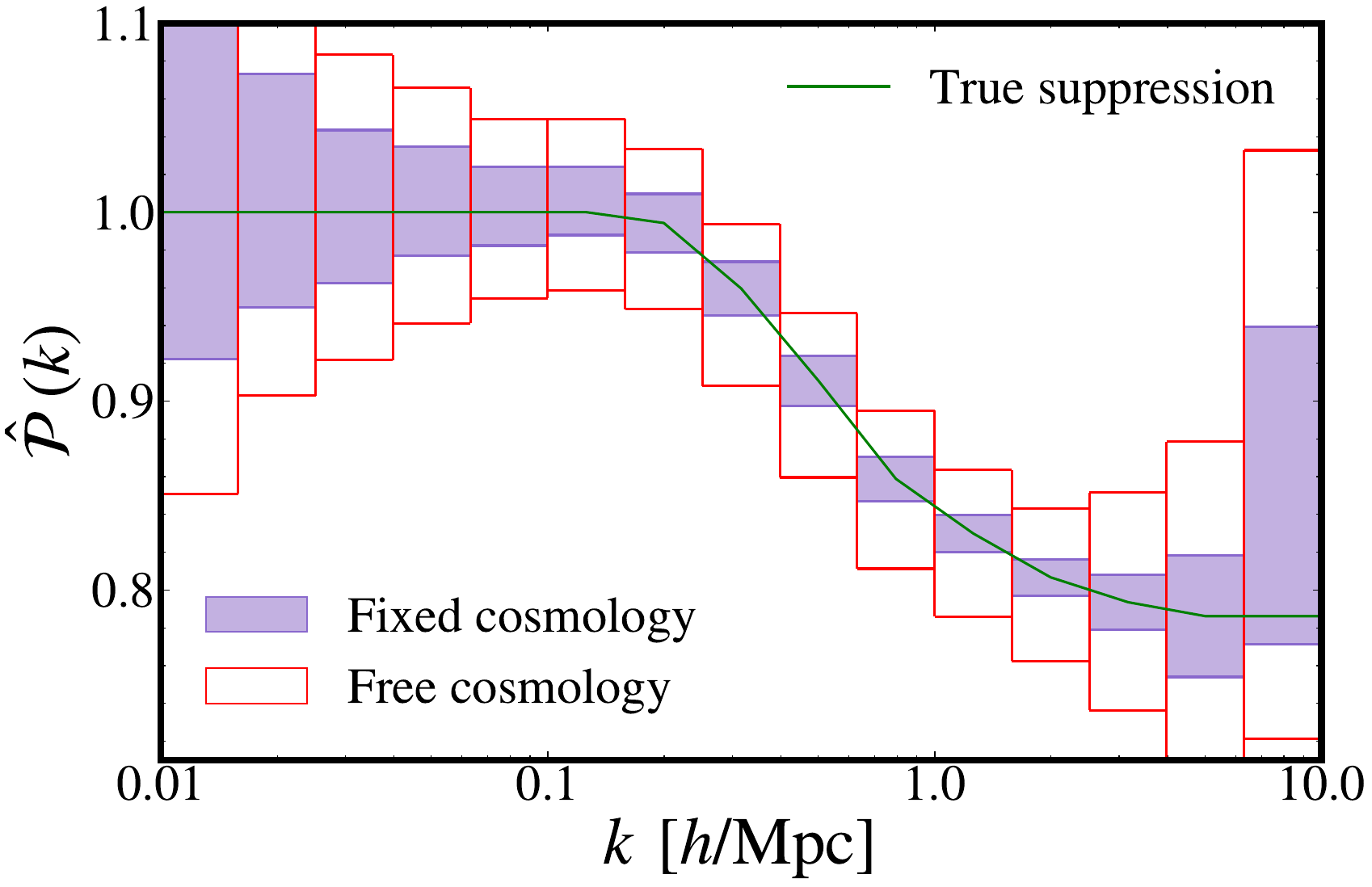} 
\caption{Comparison of reconstructing the non-linear matter power spectrum for DESC Y10 simulations with $A_{\rm{mod}}=0.8$ fixing to \Planck cosmology as in Fig.~\ref{fig:PKSAMPLEY1vsY10vsDES_SUPPRESSION} (filled constraint) and allowing cosmological parameters to vary (unfilled constraint). We see that in both cases, the true suppression is recovered, shown by the green line. However, the uncertainties are inflated when the cosmological parameters are allowed to vary by approximately $2.5\times$ in the range $0.1 < k < 1.0$.}
\label{fig:free_cosmology}
\end{figure}

Typically, in analyses of cosmic shear, cosmological parameters have been allowed to vary within uninformative priors. With current surveys such as DES, KiDS, and HSC, parameters such as $h$ and the scalar spectral index $n_{\rm{s}}$ are essentially unconstrained. The parameters $\sigma_8$ and $\Omega_{\rm{m}}$ are strongly degenerate and so each individual parameter cannot be constrained. The parameter $S_8$ measures the perpendicular direction to the $\sigma_8-\Omega_{\rm{m}}$ degeneracy and is quite well constrained with current cosmic shear surveys.

Up until now we have reconstructed the matter power spectrum, with cosmological parameters fixed to those of the best fit base \Planck \LCDM cosmology. As noted in Sect.~\ref{sec:data}, since we solve for the the power spectrum, its shape is not required to match the spectrum expected in the \Planck cosmology. The only effect of fixing to the \Planck \LCDM parameters is to constrain the expansion rate of the background cosmology and to fix the linear growth rate.

In any practical application to real data, however, the coordinate and angular diameter distance-redshift relations can be constrained in a model-independent way by using the Type 1a supernova magnitude-redshift relation together with BAO measurements to constrain $H(z)$. Furthermore, if one is willing to assume that dark energy can be approximated as uniform on scales much smaller than the Hubble radius, the form of $H(z)$ can be used to constrain the linear growth rate. There is therefore no need to rely on \Planck\ or other CMB measurements to constrain the expansion history.

Nevertheless, it is useful to analyse our simulations allowing cosmological parameters to vary, without imposing any constraints from other data, as in traditional analyses of cosmic shear. We have therefore performed power spectrum reconstructions where, in addition to solving for the power spectrum reconstruction parameters, we vary the cosmological parameters $\Omega_{\rm{m}}$, $\Omega_{\rm{b}}$, $h$, $n_{\rm s}$ and $A_{s}$, fixing $w_{0}=-1$ adopting the priors listed in \citet{DES:2021}.

The reconstructed power spectrum with free cosmology for DESC Y10 is compared to the fixed cosmology reconstruction in  Fig.~\ref{fig:free_cosmology}. Allowing cosmological parameters to vary increases the errors by factors of between $\sim 2-4$, but the $A_{\rm mod} = 0.8$ suppression assumed in the simulations is recovered at many standard deviations. We regard the free cosmology example shown in Fig.~\ref{fig:free_cosmology} to be overly pessimistic since by the time that DESC Y10 data become available there will be a huge volume of supplementary data that can be used to constrain the expansion history.

\section{Discussion and Outlook} 
\label{sec:conclusion}
\begin{figure*}
    \begin{minipage}{0.49\textwidth}
    	\includegraphics[width=1.\columnwidth]{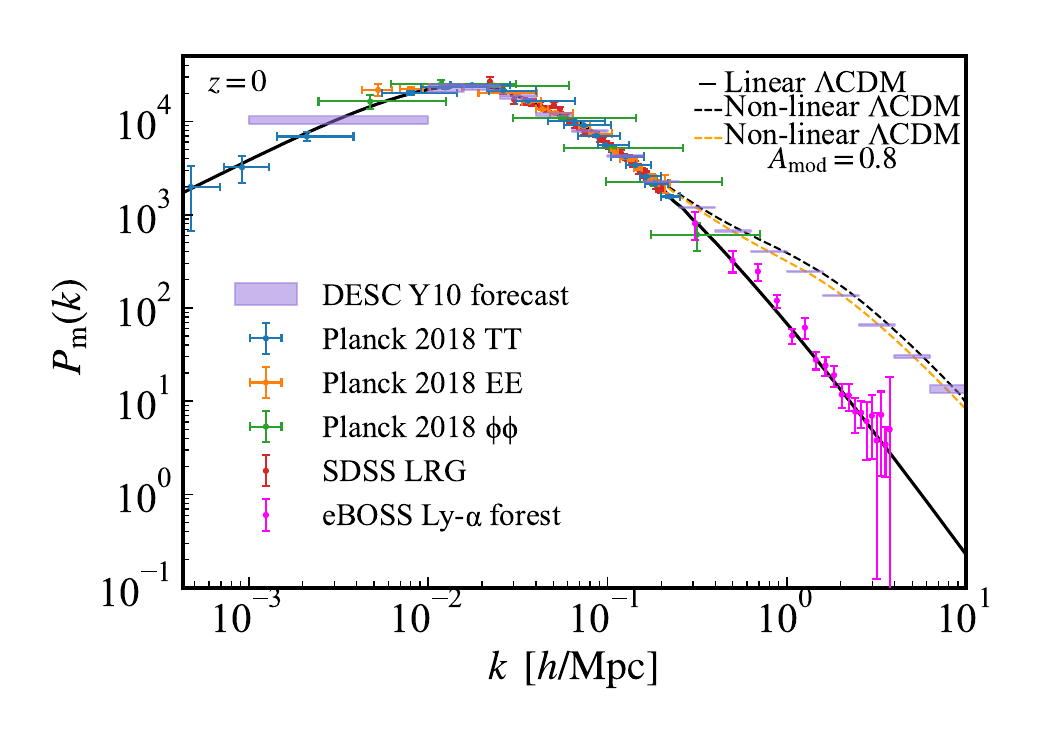} 
    \end{minipage}
    \begin{minipage}{0.49\textwidth}
        \includegraphics[width=1.\columnwidth]{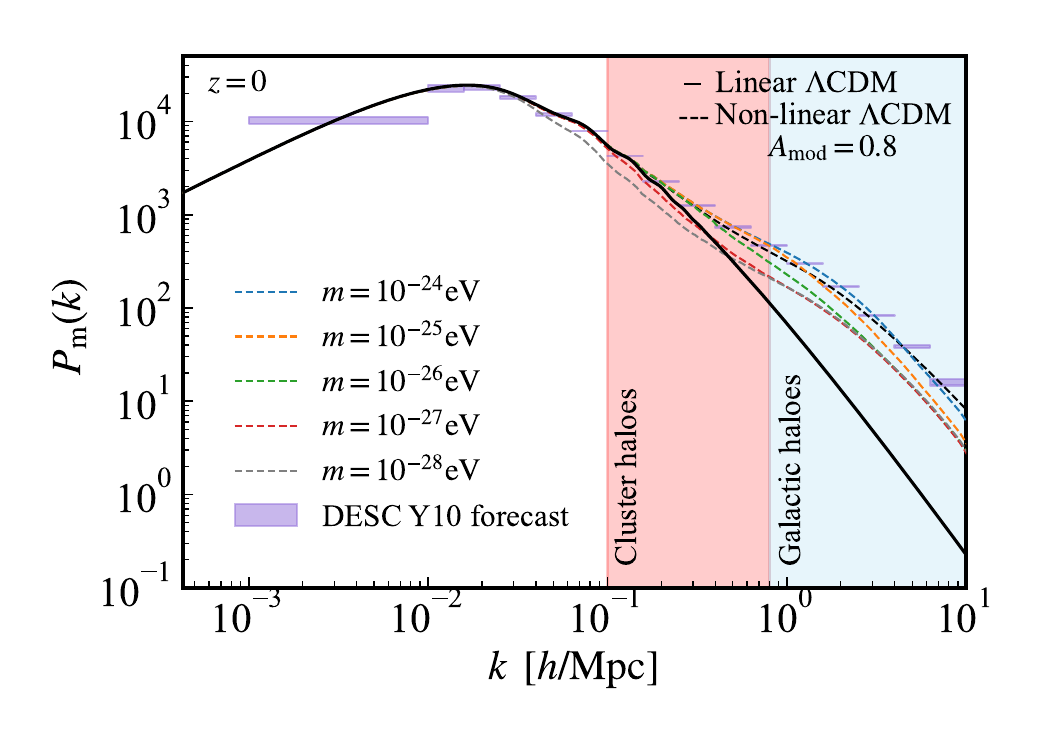} 
    \end{minipage}
\caption{The predicted linear and non-linear matter power spectrum for a dark matter only \Planck \LCDM cosmology (black) with forecasted DESC Y10 constraints (purple boxes). The non-linear \LCDM spectrum (dashed black) is predicted using Euclid Emulator. The left panel compares these constraints with others from the literature. We report constraints from \Planck \citep{Params:2018}, the Sloan Digital Sky Survey \citep{SDSS_LRG} and from the Lyman-$\alpha$ forest, as inferred from quasars \citep{eBOSSDR14}. DESC Y10 WL can place competitive constraints on the non-linear matter power spectrum. The right panel shows our constraints compared to predictions for the non-linear spectrum using a mixed dark matter model, composed of cold dark matter and ultralight axion-like particles, presented in \citet{vogt}. We show a range of ultralight axion masses for the case when axions contribute 10\% of the matter density. Our results demonstrate that future cosmic shear constraints will be sensitive enough to distinguish between these dark matter models. Shaded regions approximately delineate the wavenumber ranges to which clusters and galaxy haloes contribute (as given in Fig. 1 of \citet{White_DM_HALOES}.)}
\label{fig:Pk_total_plots}
\end{figure*}

Cosmic shear is a unique cosmological probe. It offers a window to low redshifts ($z \simlt 2$) and to non-linear scales. In addition, it tests the propogation of photons and can therefore constrain theories of gravity that display `gravitational slip' \citep[see e.g.][]{Daniel:2008, Bertschinger:2011, Simpson:2013}. Cosmic shear is therefore complementary to other cosmological probes such as the CMB, galaxy clustering and redshift-space distortions.

In most analyses of cosmic shear surveys, the primary goal has been to test whether structure formation at low redshifts is consistent with the predictions inferred from observations of the CMB. To do so requires accurate modelling of non-linear scales, to ensure that linear theory parameters such as $S_8$ are unbiased. More importantly, the standard approach mixes information from a range of scales and misses the opportunity to investigate non-linear physics.

We have investigated a number of technical issues:
\begin{enumerate}
    \item In most of our results, we assumed a \Planck \LCDM cosmology which fixes the background expansion history and the growth rate of fluctuations. However, if we adopt non-informative wide priors on cosmological parameters ignoring all other cosmological data (surely overly conservative), we find that we are able to reconstruct unbiased estimates of the non-linear power spectrum accurately, though with larger errors than in the case of fixed \Planck cosmology (see Fig.~\ref{fig:free_cosmology}).
    \item We have presented forecasts for our method using the ten-year DESC cosmic shear survey \citep{DESC_requirements}. Already, forecasts with the one-year DESC cosmic shear show a major improvement compared to current constraints from DES Y3, and should allow accurate power spectrum reconstruction (Fig.~\ref{fig:PKSAMPLEY1vsY10vsDES_SUPPRESSION}). For example, it should be possible with one or two years of data with DESC to distinguish between different models of baryonic feedback (left hand panel of Fig.~\ref{fig:Baryons_IA_Pk}, provided systematic errors are as small as stated in the DESC SRD (Fig.~\ref{fig:systematics_comparison}).
    \item Using the NLA and TATT models as a guide, IA should not lead to significant biases in the power spectrum reconstructions on scales $k \simgt 5 h{\rm / Mpc}$, where the power spectrum suppression from baryonic feedback processes is likely to exceed 10\% (right panel of Fig.~\ref{fig:Baryons_IA_Pk}). However, further work is required to 
    assess the impact of IA on larger scales, that might bias the linear theory value of $S_8$ at the percent level.
\end{enumerate}

The left hand panel of Fig.~\ref{fig:Pk_total_plots} summarises constraints on the matter power spectrum at $z=0$ from linear CMB scales to highly non-linear scales \citep[adapted from][]{PKZ_compilation}. The solid line shows the linear theory spectrum of the base \Planck \LCDM cosmology extrapolated to $z=0$, while the points labelled \Planck 2018 TT, EE, and $\phi\phi$ shows constraints inferred from the \Planck TT, EE and lensing power spectra respectively \citep{Params:2018,Plensing:2020}. \citet{PKZ_compilation} also plot constraints derived from the Sloan Digital Sky Survey galaxy halo power spectrum analysis of \cite{SDSS_LRG} and from BOSS quasar spectra \citep{eBOSSDR14} in the redshift range $2.2-4.6$ as analysed by \citet{Chabanier:2019b}. This plot\footnote{Based on data provided by \citet{PKZ_compilation}, \texttt{https://github.com/marius311/mpkcompilation}.}, which is schematic, serves to illustrate the range of scales sampled by different experiments and their relative precision. The purple boxes show the DESC Y10 reconstruction from Fig.~\ref{fig:PKSAMPLEY1vsY10vsDES_SUPPRESSION}, extrapolated to $z=0$. The black dotted line labelled shows the non-linear matter power spectrum at $z=0$ computed using the Euclid Emulator while the orange dashed line shows the suppressed spectrum with $A_{\rm mod} = 0.8$ that was used to generate the DESC Y10 data vector. As Fig.~\ref{fig:PKSAMPLEY1vsY10vsDES_SUPPRESSION} shows, the differences between these two spectra as distinguishable by DESC Y10 at many standard deviations.

The right panel of Fig.~\ref{fig:Pk_total_plots} highlights how it may be possible to distinguish between dark matter models using DESC Y10 weak lensing. The plot predicted non-linear power spectra for the specific case of a light axion particle contributing 10\% of the matter energy density, considering various values of the axion mass, computed using \texttt{axionHMCODE} \citep{vogt}. However, we note that for these predictions, baryon feedback effects have been ignored. To make inferences on the nature of dark matter from weak lensing measurements will require disentangling baryonic feedback from the effects caused by exotic dark matter. Additional data, such as thermal and kinetic Sunyaev-Zel'dovich measurements should provide valuable constraints on the impact of baryonic feedback on weak lensing \citep[e.g.][]{troester:2021, to2024, Bigwood:2024}.

In this paper we have concentrated on recovering information of the matter power spectrum on non-linear scales. Returning to the $S_8$ tension on linear scales, constraints from forthcoming galaxy surveys \citep[e.g.][]{DESI:2016} should prove decisive. In particular, redshift space distortion measurements and CMB-galaxy cross correlations \citep[e.g.][]{ACTxUNWISE} should measure the growth rate of fluctuations accurately over the same redshift range that DESC weak lensing tests. Together, these new experiments will build a complete view of the low-redshift Universe.

\section*{Acknowledgements}
Calvin Preston is supported by a Science and Technology Facilities Council studentship. Alexandra Amon was supported by a Kavli Fellowship for part of the duration of this work. George Efstathiou is grateful for the award of a Leverhulme Trust Emeritus Fellowship and for the support of the Institute of Astronomy and the Kavli Institute for Cosmology Cambridge. 
\section*{Data availability}
The data generated in this research was made following the DESC SRD \citep{DESC_requirements}.

\bibliographystyle{mnras} 
\bibliography{powerPKZ}

\begin{thebibliography}{}
\makeatletter
\relax
\def\mn@urlcharsother{\let\do\@makeother \do\$\do\&\do\#\do\^\do\_\do\%\do\~}
\def\mn@doi{\begingroup\mn@urlcharsother \@ifnextchar [ {\mn@doi@}
  {\mn@doi@[]}}
\def\mn@doi@[#1]#2{\def\@tempa{#1}\ifx\@tempa\@empty \href
  {http://dx.doi.org/#2} {doi:#2}\else \href {http://dx.doi.org/#2} {#1}\fi
  \endgroup}
\def\mn@eprint#1#2{\mn@eprint@#1:#2::\@nil}
\def\mn@eprint@arXiv#1{\href {http://arxiv.org/abs/#1} {{\tt arXiv:#1}}}
\def\mn@eprint@dblp#1{\href {http://dblp.uni-trier.de/rec/bibtex/#1.xml}
  {dblp:#1}}
\def\mn@eprint@#1:#2:#3:#4\@nil{\def\@tempa {#1}\def\@tempb {#2}\def\@tempc
  {#3}\ifx \@tempc \@empty \let \@tempc \@tempb \let \@tempb \@tempa \fi \ifx
  \@tempb \@empty \def\@tempb {arXiv}\fi \@ifundefined
  {mn@eprint@\@tempb}{\@tempb:\@tempc}{\expandafter \expandafter \csname
  mn@eprint@\@tempb\endcsname \expandafter{\@tempc}}}

\bibitem[\protect\citeauthoryear{{Abbott} et~al.,}{{Abbott}
  et~al.}{2021}]{DES:2021}
{Abbott} T.~M.~C.,  et~al., 2021, \mn@doi [\apjs] {10.3847/1538-4365/ac00b3},
  \href {https://ui.adsabs.harvard.edu/abs/2021ApJS..255...20A} {255, 20}

\bibitem[\protect\citeauthoryear{{Abdalla} et~al.,}{{Abdalla}
  et~al.}{2022}]{SNOWMASS_tensionsreview}
{Abdalla} E.,  et~al., 2022, \mn@doi [Journal of High Energy Astrophysics]
  {10.1016/j.jheap.2022.04.002}, \href
  {https://ui.adsabs.harvard.edu/abs/2022JHEAp..34...49A} {34, 49}

\bibitem[\protect\citeauthoryear{{Abolfathi} et~al.,}{{Abolfathi}
  et~al.}{2018}]{eBOSSDR14}
{Abolfathi} B.,  et~al., 2018, \mn@doi [\apjs] {10.3847/1538-4365/aa9e8a},
  \href {https://ui.adsabs.harvard.edu/abs/2018ApJS..235...42A} {235, 42}

\bibitem[\protect\citeauthoryear{{Alam} et~al.,}{{Alam}
  et~al.}{2021}]{Alam:2021a}
{Alam} S.,  et~al., 2021, \mn@doi [\prd] {10.1103/PhysRevD.103.083533}, \href
  {https://ui.adsabs.harvard.edu/abs/2021PhRvD.103h3533A} {103, 083533}

\bibitem[\protect\citeauthoryear{{Amon} \& {Efstathiou}}{{Amon} \&
  {Efstathiou}}{2022}]{AAGPE2022}
{Amon} A.,  {Efstathiou} G.,  2022, \mn@doi [\mnras] {10.1093/mnras/stac2429},
  \href {https://ui.adsabs.harvard.edu/abs/2022MNRAS.516.5355A} {516, 5355}

\bibitem[\protect\citeauthoryear{{Amon} et~al.,}{{Amon}
  et~al.}{2023}]{amon:2022}
{Amon} A.,  et~al., 2023, \mn@doi [\mnras] {10.1093/mnras/stac2938}, \href
  {https://ui.adsabs.harvard.edu/abs/2023MNRAS.518..477A} {518, 477}

\bibitem[\protect\citeauthoryear{{Angulo} \& {Hahn}}{{Angulo} \&
  {Hahn}}{2022}]{Dark_matter_nbody_review}
{Angulo} R.~E.,  {Hahn} O.,  2022, \mn@doi [Living Reviews in Computational
  Astrophysics] {10.1007/s41115-021-00013-z}, \href
  {https://ui.adsabs.harvard.edu/abs/2022LRCA....8....1A} {8, 1}

\bibitem[\protect\citeauthoryear{{Asgari} et~al.,}{{Asgari}
  et~al.}{2020}]{Asgari:2020}
{Asgari} M.,  et~al., 2020, \mn@doi [\aap] {10.1051/0004-6361/201936512}, \href
  {https://ui.adsabs.harvard.edu/abs/2020A&A...634A.127A} {634, A127}

\bibitem[\protect\citeauthoryear{{Asgari} et~al.,}{{Asgari}
  et~al.}{2021}]{Asgari:2021}
{Asgari} M.,  et~al., 2021, \mn@doi [\aap] {10.1051/0004-6361/202039070}, \href
  {https://ui.adsabs.harvard.edu/abs/2021A&A...645A.104A} {645, A104}

\bibitem[\protect\citeauthoryear{Bakx, Kurita, Chisari, Vlah  \& Schmidt}{Bakx
  et~al.}{2023}]{Bakx2023}
Bakx T.,  Kurita T.,  Chisari N.~E.,  Vlah Z.,   Schmidt F.,  2023, \mn@doi
  [JCAP] {10.1088/1475-7516/2023/10/005}, 10, 005

\bibitem[\protect\citeauthoryear{{Bechtol} et~al.,}{{Bechtol}
  et~al.}{2022}]{White_DM_HALOES}
{Bechtol} K.,  et~al., 2022, \mn@doi [arXiv e-prints]
  {10.48550/arXiv.2203.07354}, \href
  {https://ui.adsabs.harvard.edu/abs/2022arXiv220307354B} {p. arXiv:2203.07354}

\bibitem[\protect\citeauthoryear{{Bertschinger}}{{Bertschinger}}{2011}]{Bertschinger:2011}
{Bertschinger} E.,  2011, \mn@doi [Philosophical Transactions of the Royal
  Society of London Series A] {10.1098/rsta.2011.0369}, \href
  {https://ui.adsabs.harvard.edu/abs/2011RSPTA.369.4947B} {369, 4947}

\bibitem[\protect\citeauthoryear{{Bigwood} et~al.,}{{Bigwood}
  et~al.}{2024}]{Bigwood:2024}
{Bigwood} L.,  et~al., 2024, \mn@doi [arXiv e-prints]
  {10.48550/arXiv.2404.06098}, \href
  {https://ui.adsabs.harvard.edu/abs/2024arXiv240406098B} {p. arXiv:2404.06098}

\bibitem[\protect\citeauthoryear{{Blazek}, {MacCrann}, {Troxel}  \&
  {Fang}}{{Blazek} et~al.}{2019}]{BlazekTATT}
{Blazek} J.~A.,  {MacCrann} N.,  {Troxel} M.~A.,   {Fang} X.,  2019, \mn@doi
  [\prd] {10.1103/PhysRevD.100.103506}, \href
  {https://ui.adsabs.harvard.edu/abs/2019PhRvD.100j3506B} {100, 103506}

\bibitem[\protect\citeauthoryear{{Bridle} \& {King}}{{Bridle} \&
  {King}}{2007}]{Bridle_KingNLA}
{Bridle} S.,  {King} L.,  2007, \mn@doi [New Journal of Physics]
  {10.1088/1367-2630/9/12/444}, \href
  {https://ui.adsabs.harvard.edu/abs/2007NJPh....9..444B} {9, 444}

\bibitem[\protect\citeauthoryear{{Bridle}, {Lewis}, {Weller}  \&
  {Efstathiou}}{{Bridle} et~al.}{2003}]{Bridle:2003}
{Bridle} S.~L.,  {Lewis} A.~M.,  {Weller} J.,   {Efstathiou} G.,  2003, \mn@doi
  [\mnras] {10.1046/j.1365-8711.2003.06807.x}, \href
  {https://ui.adsabs.harvard.edu/abs/2003MNRAS.342L..72B} {342, L72}

\bibitem[\protect\citeauthoryear{{Brout} et~al.,}{{Brout}
  et~al.}{2022}]{Brout:2021}
{Brout} D.,  et~al., 2022, \mn@doi [\apj] {10.3847/1538-4357/ac8bcc}, \href
  {https://ui.adsabs.harvard.edu/abs/2022ApJ...938..111B} {938, 111}

\bibitem[\protect\citeauthoryear{{Brown}, {Taylor}, {Hambly}  \& {Dye}}{{Brown}
  et~al.}{2002}]{BrownIA}
{Brown} M.~L.,  {Taylor} A.~N.,  {Hambly} N.~C.,   {Dye} S.,  2002, \mn@doi
  [\mnras] {10.1046/j.1365-8711.2002.05354.x}, \href
  {https://ui.adsabs.harvard.edu/abs/2002MNRAS.333..501B} {333, 501}

\bibitem[\protect\citeauthoryear{{Catelan}, {Kamionkowski}  \&
  {Blandford}}{{Catelan} et~al.}{2001}]{Catelan:2001}
{Catelan} P.,  {Kamionkowski} M.,   {Blandford} R.~D.,  2001, \mn@doi [\mnras]
  {10.1046/j.1365-8711.2001.04105.x}, \href
  {https://ui.adsabs.harvard.edu/abs/2001MNRAS.320L...7C} {320, L7}

\bibitem[\protect\citeauthoryear{{Chabanier}, {Millea}  \&
  {Palanque-Delabrouille}}{{Chabanier} et~al.}{2019a}]{PKZ_compilation}
{Chabanier} S.,  {Millea} M.,   {Palanque-Delabrouille} N.,  2019a, \mn@doi
  [\mnras] {10.1093/mnras/stz2310}, \href
  {https://ui.adsabs.harvard.edu/abs/2019MNRAS.489.2247C} {489, 2247}

\bibitem[\protect\citeauthoryear{{Chabanier} et~al.,}{{Chabanier}
  et~al.}{2019b}]{Chabanier:2019b}
{Chabanier} S.,  et~al., 2019b, \mn@doi [\jcap]
  {10.1088/1475-7516/2019/07/017}, \href
  {https://ui.adsabs.harvard.edu/abs/2019JCAP...07..017C} {2019, 017}

\bibitem[\protect\citeauthoryear{{Chisari} et~al.,}{{Chisari}
  et~al.}{2019}]{Chisari_2019}
{Chisari} N.~E.,  et~al., 2019, \mn@doi [The Open Journal of Astrophysics]
  {10.21105/astro.1905.06082}, \href
  {https://ui.adsabs.harvard.edu/abs/2019OJAp....2E...4C} {2, 4}

\bibitem[\protect\citeauthoryear{Collaboration}{Collaboration}{2020}]{Planck2018Inflation}
Collaboration P.,  2020, \mn@doi [Astronomy and Astrophysics]
  {10.1051/0004-6361/201833887}, 641, A10

\bibitem[\protect\citeauthoryear{{Crittenden}, {Natarajan}, {Pen}  \&
  {Theuns}}{{Crittenden} et~al.}{2001}]{Crittenden}
{Crittenden} R.~G.,  {Natarajan} P.,  {Pen} U.-L.,   {Theuns} T.,  2001,
  \mn@doi [\apj] {10.1086/322370}, \href
  {https://ui.adsabs.harvard.edu/abs/2001ApJ...559..552C} {559, 552}

\bibitem[\protect\citeauthoryear{{DES} \& {KiDS Collaborations}}{{DES} \& {KiDS
  Collaborations}}{2023}]{KiDSDES}
{DES} {KiDS Collaborations} 2023, The Open Journal of Astrophysics

\bibitem[\protect\citeauthoryear{{DESI Collaboration} et~al.,}{{DESI
  Collaboration} et~al.}{2016}]{DESI:2016}
{DESI Collaboration} et~al., 2016, arXiv e-prints, \href
  {https://ui.adsabs.harvard.edu/abs/2016arXiv161100036D} {p. arXiv:1611.00036}

\bibitem[\protect\citeauthoryear{{DESI collaboration} et~al.,}{{DESI
  collaboration} et~al.}{2024}]{DESI_BAO_2024}
{DESI collaboration} et~al., 2024, DESI 2024 VI: Cosmological Constraints from
  the Measurements of Baryon Acoustic Oscillations (\mn@eprint {arXiv}
  {2404.03002})

\bibitem[\protect\citeauthoryear{Dalal et~al.,}{Dalal
  et~al.}{2023}]{dalal2023hyper}
Dalal R.,  et~al., 2023, Hyper Suprime-Cam Year 3 Results: Cosmology from
  Cosmic Shear Power Spectra (\mn@eprint {arXiv} {2304.00701})

\bibitem[\protect\citeauthoryear{{Daniel}, {Caldwell}, {Cooray}  \&
  {Melchiorri}}{{Daniel} et~al.}{2008}]{Daniel:2008}
{Daniel} S.~F.,  {Caldwell} R.~R.,  {Cooray} A.,   {Melchiorri} A.,  2008,
  \mn@doi [\prd] {10.1103/PhysRevD.77.103513}, \href
  {https://ui.adsabs.harvard.edu/abs/2008PhRvD..77j3513D} {77, 103513}

\bibitem[\protect\citeauthoryear{{Doux} et~al.,}{{Doux}
  et~al.}{2022}]{DouxDESHarmonic}
{Doux} C.,  et~al., 2022, \mn@doi [\mnras] {10.1093/mnras/stac1826}, \href
  {https://ui.adsabs.harvard.edu/abs/2022MNRAS.515.1942D} {515, 1942}

\bibitem[\protect\citeauthoryear{{Dubois} et~al.,}{{Dubois}
  et~al.}{2014}]{Dubois:2014}
{Dubois} Y.,  et~al., 2014, \mn@doi [\mnras] {10.1093/mnras/stu1227}, \href
  {https://ui.adsabs.harvard.edu/abs/2014MNRAS.444.1453D} {444, 1453}

\bibitem[\protect\citeauthoryear{{Efstathiou} \& {Gratton}}{{Efstathiou} \&
  {Gratton}}{2020}]{Efstathiou:2020}
{Efstathiou} G.,  {Gratton} S.,  2020, \mn@doi [\mnras]
  {10.1093/mnrasl/slaa093}, \href
  {https://ui.adsabs.harvard.edu/abs/2020MNRAS.496L..91E} {496, L91}

\bibitem[\protect\citeauthoryear{Elbers et~al.,}{Elbers
  et~al.}{2024}]{Elbers:2024}
Elbers W.,  et~al., 2024, The FLAMINGO project: the coupling between baryonic
  feedback and cosmology in light of the $S_8$ tension (\mn@eprint {arXiv}
  {2403.12967})

\bibitem[\protect\citeauthoryear{{Euclid Collaboration} et~al.,}{{Euclid
  Collaboration} et~al.}{2020}]{EuclidForecast}
{Euclid Collaboration} et~al., 2020, \mn@doi [\aap]
  {10.1051/0004-6361/202038071}, \href
  {https://ui.adsabs.harvard.edu/abs/2020A&A...642A.191E} {642, A191}

\bibitem[\protect\citeauthoryear{{Euclid Collaboration} et~al.,}{{Euclid
  Collaboration} et~al.}{2021}]{EuclidEmulator}
{Euclid Collaboration} et~al., 2021, \mn@doi [\mnras] {10.1093/mnras/stab1366},
  \href {https://ui.adsabs.harvard.edu/abs/2021MNRAS.505.2840E} {505, 2840}

\bibitem[\protect\citeauthoryear{Farren et~al.,}{Farren
  et~al.}{2023}]{ACTxUNWISE}
Farren G.~S.,  et~al., 2023, The Atacama Cosmology Telescope: Cosmology from
  cross-correlations of unWISE galaxies and ACT DR6 CMB lensing (\mn@eprint
  {arXiv} {2309.05659})

\bibitem[\protect\citeauthoryear{Feroz, Hobson  \& Bridges}{Feroz
  et~al.}{2009}]{Feroz_2009}
Feroz F.,  Hobson M.~P.,   Bridges M.,  2009, \mn@doi [Monthly Notices of the
  Royal Astronomical Society] {10.1111/j.1365-2966.2009.14548.x}, 398, 1601

\bibitem[\protect\citeauthoryear{Fortuna, Hoekstra, Joachimi, Johnston,
  Chisari, Georgiou  \& Mahony}{Fortuna et~al.}{2020}]{Fortuna2020}
Fortuna M.~C.,  Hoekstra H.,  Joachimi B.,  Johnston H.,  Chisari N.~E.,
  Georgiou C.,   Mahony C.,  2020, \mn@doi [Monthly Notices of the Royal
  Astronomical Society] {10.1093/mnras/staa3802}, 501, 2983–3002

\bibitem[\protect\citeauthoryear{{Fortuna} et~al.,}{{Fortuna}
  et~al.}{2021}]{KiDSIA}
{Fortuna} M.~C.,  et~al., 2021, \mn@doi [\aap] {10.1051/0004-6361/202140706},
  \href {https://ui.adsabs.harvard.edu/abs/2021A&A...654A..76F} {654, A76}

\bibitem[\protect\citeauthoryear{{Handley}, {Lasenby}, {Peiris}  \&
  {Hobson}}{{Handley} et~al.}{2019}]{Handley:2019}
{Handley} W.~J.,  {Lasenby} A.~N.,  {Peiris} H.~V.,   {Hobson} M.~P.,  2019,
  \mn@doi [\prd] {10.1103/PhysRevD.100.103511}, \href
  {https://ui.adsabs.harvard.edu/abs/2019PhRvD.100j3511H} {100, 103511}

\bibitem[\protect\citeauthoryear{{Heymans} et~al.,}{{Heymans}
  et~al.}{2013}]{Heymans:2013}
{Heymans} C.,  et~al., 2013, \mn@doi [\mnras] {10.1093/mnras/stt601}, \href
  {http://adsabs.harvard.edu/abs/2013MNRAS.432.2433H} {432, 2433}

\bibitem[\protect\citeauthoryear{{Hilbert}, {Xu}, {Schneider}, {Springel},
  {Vogelsberger}  \& {Hernquist}}{{Hilbert} et~al.}{2017}]{Hilbert:2017}
{Hilbert} S.,  {Xu} D.,  {Schneider} P.,  {Springel} V.,  {Vogelsberger} M.,
  {Hernquist} L.,  2017, \mn@doi [\mnras] {10.1093/mnras/stx482}, \href
  {https://ui.adsabs.harvard.edu/abs/2017MNRAS.468..790H} {468, 790}

\bibitem[\protect\citeauthoryear{{Hildebrandt} et~al.,}{{Hildebrandt}
  et~al.}{2017}]{Hildebrandt:2017}
{Hildebrandt} H.,  et~al., 2017, \mn@doi [\mnras] {10.1093/mnras/stw2805},
  \href {https://ui.adsabs.harvard.edu/abs/2017MNRAS.465.1454H} {465, 1454}

\bibitem[\protect\citeauthoryear{{Hirata} \& {Seljak}}{{Hirata} \&
  {Seljak}}{2004}]{Hirata&Seljak}
{Hirata} C.~M.,  {Seljak} U.,  2004, \mn@doi [\prd]
  {10.1103/PhysRevD.70.063526}, \href
  {https://ui.adsabs.harvard.edu/abs/2004PhRvD..70f3526H} {70, 063526}

\bibitem[\protect\citeauthoryear{{Joachimi} et~al.,}{{Joachimi}
  et~al.}{2015}]{Joachimi:2015}
{Joachimi} B.,  et~al., 2015, \mn@doi [\ssr] {10.1007/s11214-015-0177-4}, \href
  {https://ui.adsabs.harvard.edu/abs/2015SSRv..193....1J} {193, 1}

\bibitem[\protect\citeauthoryear{Johnston et~al.,}{Johnston
  et~al.}{2019}]{Johnston_2019}
Johnston H.,  et~al., 2019, \mn@doi [Astronomy &amp; Astrophysics]
  {10.1051/0004-6361/201834714}, 624, A30

\bibitem[\protect\citeauthoryear{{Joudaki} et~al.,}{{Joudaki}
  et~al.}{2017}]{Joudaki:2017cfht}
{Joudaki} S.,  et~al., 2017, \mn@doi [\mnras] {10.1093/mnras/stw2665}, \href
  {https://ui.adsabs.harvard.edu/abs/2017MNRAS.465.2033J} {465, 2033}

\bibitem[\protect\citeauthoryear{{LSST Science Collaboration} et~al.,}{{LSST
  Science Collaboration} et~al.}{2009}]{LSSTScience}
{LSST Science Collaboration} et~al., 2009, \mn@doi [arXiv e-prints]
  {10.48550/arXiv.0912.0201}, \href
  {https://ui.adsabs.harvard.edu/abs/2009arXiv0912.0201L} {p. arXiv:0912.0201}

\bibitem[\protect\citeauthoryear{{Lagu{\"e}}, {Schwabe}, {Hlo{\v{z}}ek},
  {Marsh}  \& {Rogers}}{{Lagu{\"e}} et~al.}{2024}]{Lague}
{Lagu{\"e}} A.,  {Schwabe} B.,  {Hlo{\v{z}}ek} R.,  {Marsh} D. J.~E.,
  {Rogers} K.~K.,  2024, \mn@doi [\prd] {10.1103/PhysRevD.109.043507}, \href
  {https://ui.adsabs.harvard.edu/abs/2024PhRvD.109d3507L} {109, 043507}

\bibitem[\protect\citeauthoryear{{Lamman}, {Tsaprazi}, {Shi},
  {{\v{S}}ar{\v{c}}evi{\'c}}, {Pyne}, {Legnani}  \& {Ferreira}}{{Lamman}
  et~al.}{2024}]{Lamman:2024}
{Lamman} C.,  {Tsaprazi} E.,  {Shi} J.,  {{\v{S}}ar{\v{c}}evi{\'c}} N.~N.,
  {Pyne} S.,  {Legnani} E.,   {Ferreira} T.,  2024, \mn@doi [The Open Journal
  of Astrophysics] {10.21105/astro.2309.08605}, \href
  {https://ui.adsabs.harvard.edu/abs/2024OJAp....7E..14L} {7, 14}

\bibitem[\protect\citeauthoryear{{Lewis} \& {Challinor}}{{Lewis} \&
  {Challinor}}{2011}]{CAMB}
{Lewis} A.,  {Challinor} A.,  2011, {CAMB: Code for Anisotropies in the
  Microwave Background}, Astrophysics Source Code Library, record ascl:1102.026
  (\mn@eprint {ascl} {1102.026})

\bibitem[\protect\citeauthoryear{Li et~al.,}{Li et~al.}{2023}]{li2023hyper}
Li X.,  et~al., 2023, Hyper Suprime-Cam Year 3 Results: Cosmology from Cosmic
  Shear Two-point Correlation Functions (\mn@eprint {arXiv} {2304.00702})

\bibitem[\protect\citeauthoryear{{Liu}, {Bird}, {Zorrilla Matilla}, {Hill},
  {Haiman}, {Madhavacheril}, {Petri}  \& {Spergel}}{{Liu}
  et~al.}{2018}]{Liu:2018}
{Liu} J.,  {Bird} S.,  {Zorrilla Matilla} J.~M.,  {Hill} J.~C.,  {Haiman} Z.,
  {Madhavacheril} M.~S.,  {Petri} A.,   {Spergel} D.~N.,  2018, \mn@doi [\jcap]
  {10.1088/1475-7516/2018/03/049}, \href
  {https://ui.adsabs.harvard.edu/abs/2018JCAP...03..049L} {2018, 049}

\bibitem[\protect\citeauthoryear{{Madhavacheril} et~al.,}{{Madhavacheril}
  et~al.}{2023}]{ACT2023}
{Madhavacheril} M.~S.,  et~al., 2023, \mn@doi [arXiv e-prints]
  {10.48550/arXiv.2304.05203}, \href
  {https://ui.adsabs.harvard.edu/abs/2023arXiv230405203M} {p. arXiv:2304.05203}

\bibitem[\protect\citeauthoryear{{Maion}, {Angulo}, {Bakx}, {Chisari}, {Kurita}
   \& {Pellejero-Ib{\'a}{\~n}ez}}{{Maion} et~al.}{2023}]{Maion_IA_Perturbative}
{Maion} F.,  {Angulo} R.~E.,  {Bakx} T.,  {Chisari} N.~E.,  {Kurita} T.,
  {Pellejero-Ib{\'a}{\~n}ez} M.,  2023, \mn@doi [arXiv e-prints]
  {10.48550/arXiv.2307.13754}, \href
  {https://ui.adsabs.harvard.edu/abs/2023arXiv230713754M} {p. arXiv:2307.13754}

\bibitem[\protect\citeauthoryear{{Mandelbaum}, {Hirata}, {Ishak}, {Seljak}  \&
  {Brinkmann}}{{Mandelbaum} et~al.}{2006}]{Mandelbaum2006}
{Mandelbaum} R.,  {Hirata} C.~M.,  {Ishak} M.,  {Seljak} U.,   {Brinkmann} J.,
  2006, \mn@doi [\mnras] {10.1111/j.1365-2966.2005.09946.x}, \href
  {https://ui.adsabs.harvard.edu/abs/2006MNRAS.367..611M} {367, 611}

\bibitem[\protect\citeauthoryear{{McCarthy}, {Schaye}, {Bird}  \& {Le
  Brun}}{{McCarthy} et~al.}{2017a}]{McCarthy:2017}
{McCarthy} I.~G.,  {Schaye} J.,  {Bird} S.,   {Le Brun} A. M.~C.,  2017a,
  \mn@doi [\mnras] {10.1093/mnras/stw2792}, \href
  {https://ui.adsabs.harvard.edu/abs/2017MNRAS.465.2936M} {465, 2936}

\bibitem[\protect\citeauthoryear{{McCarthy}, {Schaye}, {Bird}  \& {Le
  Brun}}{{McCarthy} et~al.}{2017b}]{BAHAMAS}
{McCarthy} I.~G.,  {Schaye} J.,  {Bird} S.,   {Le Brun} A. M.~C.,  2017b,
  \mn@doi [\mnras] {10.1093/mnras/stw2792}, \href
  {https://ui.adsabs.harvard.edu/abs/2017MNRAS.465.2936M} {465, 2936}

\bibitem[\protect\citeauthoryear{{Myles} et~al.,}{{Myles}
  et~al.}{2021}]{Myles:2021}
{Myles} J.,  et~al., 2021, \mn@doi [\mnras] {10.1093/mnras/stab1515}, \href
  {https://ui.adsabs.harvard.edu/abs/2021MNRAS.505.4249M} {505, 4249}

\bibitem[\protect\citeauthoryear{{Pakmor} et~al.,}{{Pakmor}
  et~al.}{2023}]{Pakamor:2023}
{Pakmor} R.,  et~al., 2023, \mn@doi [\mnras] {10.1093/mnras/stac3620}, \href
  {https://ui.adsabs.harvard.edu/abs/2023MNRAS.524.2539P} {524, 2539}

\bibitem[\protect\citeauthoryear{{Peiris} \& {Verde}}{{Peiris} \&
  {Verde}}{2010}]{Peiris:2010}
{Peiris} H.~V.,  {Verde} L.,  2010, \mn@doi [\prd]
  {10.1103/PhysRevD.81.021302}, \href
  {https://ui.adsabs.harvard.edu/abs/2010PhRvD..81b1302P} {81, 021302}

\bibitem[\protect\citeauthoryear{{Planck Collaboration} et~al.,}{{Planck
  Collaboration} et~al.}{2020a}]{Params:2018}
{Planck Collaboration} et~al., 2020a, \mn@doi [\aap]
  {10.1051/0004-6361/201833910}, \href
  {https://ui.adsabs.harvard.edu/abs/2020A&A...641A...6P} {641, A6}

\bibitem[\protect\citeauthoryear{{Planck Collaboration} et~al.,}{{Planck
  Collaboration} et~al.}{2020b}]{Plensing:2020}
{Planck Collaboration} et~al., 2020b, \mn@doi [\aap]
  {10.1051/0004-6361/201833886}, \href
  {https://ui.adsabs.harvard.edu/abs/2020A&A...641A...8P} {641, A8}

\bibitem[\protect\citeauthoryear{{Preston}, {Amon}  \& {Efstathiou}}{{Preston}
  et~al.}{2023}]{CPAAGPE}
{Preston} C.,  {Amon} A.,   {Efstathiou} G.,  2023, \mn@doi [\mnras]
  {10.1093/mnras/stad2573}, \href
  {https://ui.adsabs.harvard.edu/abs/2023MNRAS.525.5554P} {525, 5554}

\bibitem[\protect\citeauthoryear{{Reid} et~al.,}{{Reid}
  et~al.}{2010}]{SDSS_LRG}
{Reid} B.~A.,  et~al., 2010, \mn@doi [\mnras]
  {10.1111/j.1365-2966.2010.16276.x}, \href
  {https://ui.adsabs.harvard.edu/abs/2010MNRAS.404...60R} {404, 60}

\bibitem[\protect\citeauthoryear{{Rogers}, {Hlo{\v{z}}ek}, {Lagu{\"e}},
  {Ivanov}, {Philcox}, {Cabass}, {Akitsu}  \& {Marsh}}{{Rogers}
  et~al.}{2023a}]{Rogers:2023}
{Rogers} K.~K.,  {Hlo{\v{z}}ek} R.,  {Lagu{\"e}} A.,  {Ivanov} M.~M.,
  {Philcox} O. H.~E.,  {Cabass} G.,  {Akitsu} K.,   {Marsh} D. J.~E.,  2023a,
  \mn@doi [arXiv e-prints] {10.48550/arXiv.2301.08361}, \href
  {https://ui.adsabs.harvard.edu/abs/2023arXiv230108361R} {p. arXiv:2301.08361}

\bibitem[\protect\citeauthoryear{{Rogers}, {Hlo{\v{z}}ek}, {Lagu{\"e}},
  {Ivanov}, {Philcox}, {Cabass}, {Akitsu}  \& {Marsh}}{{Rogers}
  et~al.}{2023b}]{Rogers_S8tensionaxions}
{Rogers} K.~K.,  {Hlo{\v{z}}ek} R.,  {Lagu{\"e}} A.,  {Ivanov} M.~M.,
  {Philcox} O. H.~E.,  {Cabass} G.,  {Akitsu} K.,   {Marsh} D. J.~E.,  2023b,
  \mn@doi [\jcap] {10.1088/1475-7516/2023/06/023}, \href
  {https://ui.adsabs.harvard.edu/abs/2023JCAP...06..023R} {2023, 023}

\bibitem[\protect\citeauthoryear{{Samuroff} et~al.,}{{Samuroff}
  et~al.}{2019}]{Samuroff2019}
{Samuroff} S.,  et~al., 2019, \mn@doi [\mnras] {10.1093/mnras/stz2197}, \href
  {https://ui.adsabs.harvard.edu/abs/2019MNRAS.489.5453S} {489, 5453}

\bibitem[\protect\citeauthoryear{Samuroff et~al.,}{Samuroff
  et~al.}{2023}]{Samuroff_2023}
Samuroff S.,  et~al., 2023, \mn@doi [Monthly Notices of the Royal Astronomical
  Society] {10.1093/mnras/stad2013}, 524, 2195–2223

\bibitem[\protect\citeauthoryear{{Schaye} et~al.,}{{Schaye}
  et~al.}{2023}]{Schaye2023}
{Schaye} J.,  et~al., 2023, \mn@doi [\mnras] {10.1093/mnras/stad2419}, \href
  {https://ui.adsabs.harvard.edu/abs/2023MNRAS.526.4978S} {526, 4978}

\bibitem[\protect\citeauthoryear{{Scolnic} et~al.,}{{Scolnic}
  et~al.}{2022}]{Pantheon_plus}
{Scolnic} D.,  et~al., 2022, \mn@doi [\apj] {10.3847/1538-4357/ac8b7a}, \href
  {https://ui.adsabs.harvard.edu/abs/2022ApJ...938..113S} {938, 113}

\bibitem[\protect\citeauthoryear{{Secco}, {Samuroff}  et~al.}{{Secco}
  et~al.}{2022}]{Secco:2022}
{Secco} L.~F.,  {Samuroff} S.,   et~al., 2022, \mn@doi [\prd]
  {10.1103/PhysRevD.105.023515}, \href
  {https://ui.adsabs.harvard.edu/abs/2022PhRvD.105b3515S} {105, 023515}

\bibitem[\protect\citeauthoryear{{Simpson} et~al.,}{{Simpson}
  et~al.}{2013}]{Simpson:2013}
{Simpson} F.,  et~al., 2013, \mn@doi [\mnras] {10.1093/mnras/sts493}, \href
  {https://ui.adsabs.harvard.edu/abs/2013MNRAS.429.2249S} {429, 2249}

\bibitem[\protect\citeauthoryear{Spergel et~al.,}{Spergel et~al.}{2015}]{Roman}
Spergel D.,  et~al., 2015, Wide-Field InfrarRed Survey Telescope-Astrophysics
  Focused Telescope Assets WFIRST-AFTA 2015 Report (\mn@eprint {arXiv}
  {1503.03757})

\bibitem[\protect\citeauthoryear{{Springel} et~al.,}{{Springel}
  et~al.}{2018}]{Springel:2018}
{Springel} V.,  et~al., 2018, \mn@doi [\mnras] {10.1093/mnras/stx3304}, \href
  {https://ui.adsabs.harvard.edu/abs/2018MNRAS.475..676S} {475, 676}

\bibitem[\protect\citeauthoryear{{Tegmark} \& {Zaldarriaga}}{{Tegmark} \&
  {Zaldarriaga}}{2002}]{Tegmark_Zaldarriaga}
{Tegmark} M.,  {Zaldarriaga} M.,  2002, \mn@doi [\prd]
  {10.1103/PhysRevD.66.103508}, \href
  {https://ui.adsabs.harvard.edu/abs/2002PhRvD..66j3508T} {66, 103508}

\bibitem[\protect\citeauthoryear{{Tenneti}, {Singh}, {Mandelbaum}, {di Matteo},
  {Feng}  \& {Khandai}}{{Tenneti} et~al.}{2015}]{Tenneti:2015}
{Tenneti} A.,  {Singh} S.,  {Mandelbaum} R.,  {di Matteo} T.,  {Feng} Y.,
  {Khandai} N.,  2015, \mn@doi [\mnras] {10.1093/mnras/stv272}, \href
  {https://ui.adsabs.harvard.edu/abs/2015MNRAS.448.3522T} {448, 3522}

\bibitem[\protect\citeauthoryear{{The LSST Dark Energy Science Collaboration}
  et~al.,}{{The LSST Dark Energy Science Collaboration}
  et~al.}{2018}]{DESC_requirements}
{The LSST Dark Energy Science Collaboration} et~al., 2018, \mn@doi [arXiv
  e-prints] {10.48550/arXiv.1809.01669}, \href
  {https://ui.adsabs.harvard.edu/abs/2018arXiv180901669T} {p. arXiv:1809.01669}

\bibitem[\protect\citeauthoryear{{To}, {Pandey}, {Krause}, {Dalal},
  {Anbajagane}  \& {Weinberg}}{{To} et~al.}{2024}]{to2024}
{To} C.-H.,  {Pandey} S.,  {Krause} E.,  {Dalal} N.,  {Anbajagane} D.,
  {Weinberg} D.~H.,  2024, \mn@doi [arXiv e-prints]
  {10.48550/arXiv.2402.00110}, \href
  {https://ui.adsabs.harvard.edu/abs/2024arXiv240200110T} {p. arXiv:2402.00110}

\bibitem[\protect\citeauthoryear{{Tr{\"o}ster} et~al.,}{{Tr{\"o}ster}
  et~al.}{2022}]{troester:2021}
{Tr{\"o}ster} T.,  et~al., 2022, \mn@doi [\aap] {10.1051/0004-6361/202142197},
  \href {https://ui.adsabs.harvard.edu/abs/2022A&A...660A..27T} {660, A27}

\bibitem[\protect\citeauthoryear{{Troxel} \& {Ishak}}{{Troxel} \&
  {Ishak}}{2015}]{TroxelIA}
{Troxel} M.~A.,  {Ishak} M.,  2015, \mn@doi [\physrep]
  {10.1016/j.physrep.2014.11.001}, \href
  {https://ui.adsabs.harvard.edu/abs/2015PhR...558....1T} {558, 1}

\bibitem[\protect\citeauthoryear{{Viel}, {Markovi{\v{c}}}, {Baldi}  \&
  {Weller}}{{Viel} et~al.}{2012}]{Viel}
{Viel} M.,  {Markovi{\v{c}}} K.,  {Baldi} M.,   {Weller} J.,  2012, \mn@doi
  [\mnras] {10.1111/j.1365-2966.2011.19910.x}, \href
  {https://ui.adsabs.harvard.edu/abs/2012MNRAS.421...50V} {421, 50}

\bibitem[\protect\citeauthoryear{{Vogt}, {Marsh}  \& {Lagu{\"e}}}{{Vogt}
  et~al.}{2023}]{vogt}
{Vogt} S. M.~L.,  {Marsh} D. J.~E.,   {Lagu{\"e}} A.,  2023, \mn@doi [\prd]
  {10.1103/PhysRevD.107.063526}, \href
  {https://ui.adsabs.harvard.edu/abs/2023PhRvD.107f3526V} {107, 063526}

\bibitem[\protect\citeauthoryear{{York} et~al.,}{{York}
  et~al.}{2000}]{York:2000}
{York} D.~G.,  et~al., 2000, \mn@doi [\aj] {10.1086/301513}, \href
  {https://ui.adsabs.harvard.edu/abs/2000AJ....120.1579Y} {120, 1579}

\bibitem[\protect\citeauthoryear{{Zuntz} et~al.,}{{Zuntz}
  et~al.}{2015}]{Zuntz:2015}
{Zuntz} J.,  et~al., 2015, \mn@doi [Astronomy and Computing]
  {10.1016/j.ascom.2015.05.005}, \href
  {https://ui.adsabs.harvard.edu/abs/2015A&C....12...45Z} {12, 45}

\bibitem[\protect\citeauthoryear{{van Daalen}, {Schaye}, {Booth}  \& {Dalla
  Vecchia}}{{van Daalen} et~al.}{2011}]{vanDaalen:2011}
{van Daalen} M.~P.,  {Schaye} J.,  {Booth} C.~M.,   {Dalla Vecchia} C.,  2011,
  \mn@doi [\mnras] {10.1111/j.1365-2966.2011.18981.x}, \href
  {https://ui.adsabs.harvard.edu/abs/2011MNRAS.415.3649V} {415, 3649}

\bibitem[\protect\citeauthoryear{{van Daalen}, {McCarthy}  \& {Schaye}}{{van
  Daalen} et~al.}{2020}]{vandaalen:2020}
{van Daalen} M.~P.,  {McCarthy} I.~G.,   {Schaye} J.,  2020, \mn@doi [\mnras]
  {10.1093/mnras/stz3199}, \href
  {https://ui.adsabs.harvard.edu/abs/2020MNRAS.491.2424V} {491, 2424}

\makeatother
\end{thebibliography}
\appendix
\section{Binning of non-linear scales}
\label{sec:binning}

\begin{figure}
\centering
   \includegraphics[width=1\columnwidth]{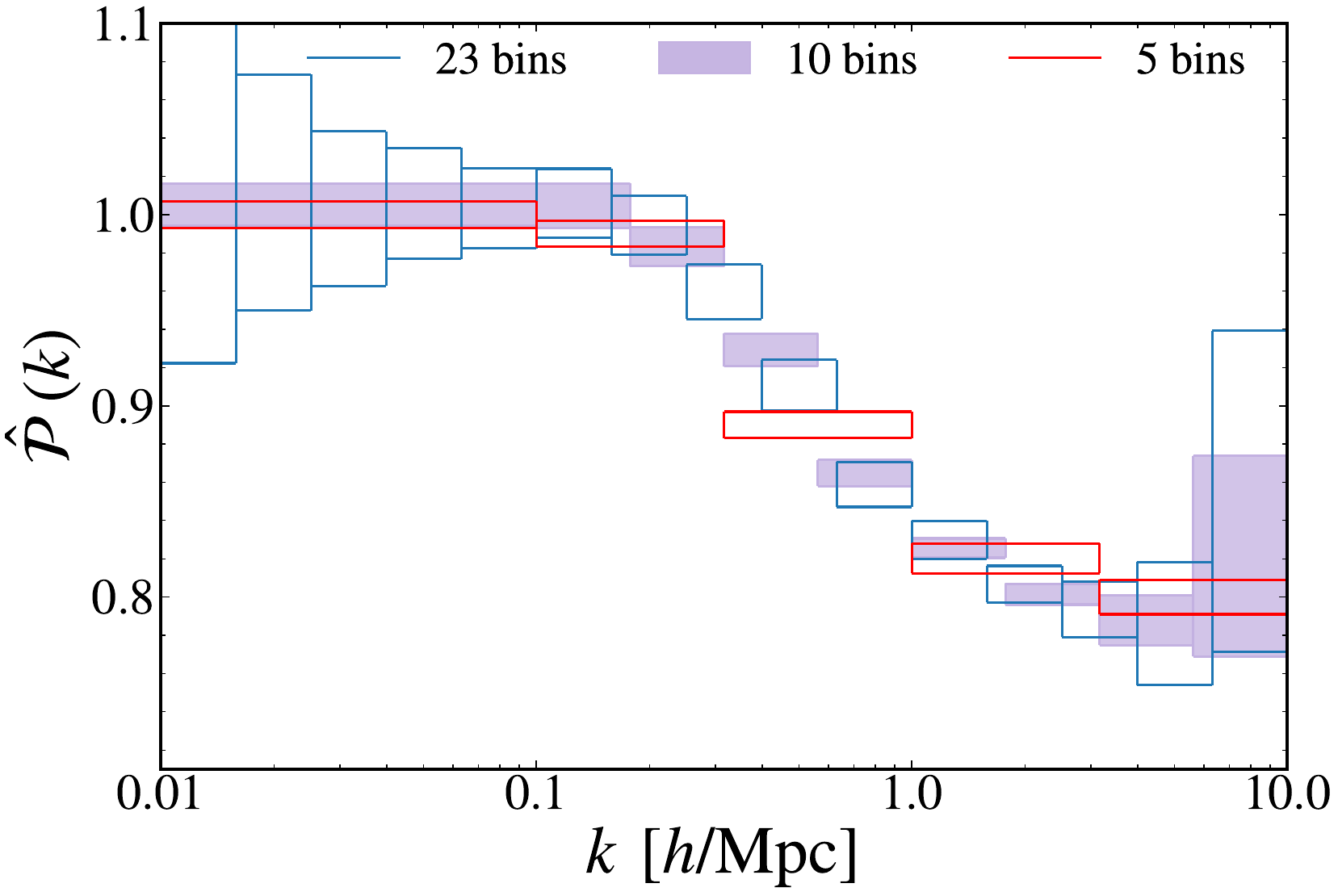} 
\caption{The fiducial suppression of the DESC Y10 data vector constrained using different numbers. The same dark matter spectrum is used here in all cases. We report results for 5, 10 and 23 bins. As binning number is increased, error bars show a flaring response, highlighting the more constraining regimes of the data vector. Too few bins loses shape information in the range $10^{-1} < k < 10$.}
\label{fig:binning_comparison}
\end{figure}

We investigate power spectrum reconstructions with different bin widths. As shown in fig.~\ref{fig:binning_comparison}, using 5, 10 and 23 bins recovers the shape of the spectrum over the wavenumber range $10^{-1}<k~[h/\rm{Mpc}]<10$, but loses resolution outside this range because of our use of open-ended bins. These issues are avoided if we adopt 23 bins evenly spaced in $\log (k)$ as reported in the main body of the paper.

Because we are reconstructing smooth spectra with no there is no gain in using finer bins.

We also investigated the effects on  reconstruction on imposing a derivative penalty. This follows previous works on constraining the shape of power spectrum reconstructions by \citet{Planck2018Inflation} (for more details see \citet{Tegmark_Zaldarriaga} and references therein). We add to the likelihood 

\begin{multline}
\label{equ:derivative_penalty}
   -2\ln \mathcal{L} = \lambda \int_{\kappa_{\rm{min}}}^{\kappa_{\rm{max}}} d\kappa \left( \frac{\partial f^{2} (\kappa)}{\partial \kappa^{2}}\right)^{2}   \\ + \alpha \int_{-\infty}^{\kappa_{\rm{min}}}     f^{2}(\kappa) \ d\kappa + \alpha \int_{\kappa_{\rm{max}}}^{\infty}   f^{2}(\kappa) \ d\kappa
\end{multline}

Where $P_{m}(k) = P_{0}(k)[1+f(\kappa)]$ has $P_{0}(k)$ as the DMO power spectrum, such that $f(\kappa)$ captures fractional deviations from the DMO spectrum, and $\kappa=\log_{10}k$. $\kappa_{\rm{min}}$ and $\kappa_{\rm{max}}$ capture the rough constraining scales of the data. $\lambda$ represents a smoothing scale of the data, whilst $\alpha$'s value is chosen to reflect the constraining power of data outside of the range $[\kappa_{\rm{min}},\kappa_{\rm{max}}]$.

We explore a range of smoothing lengths $\lambda$ for various values of $\alpha=0.001$. 

Applying a derivative penalty makes little difference to power spectrum reconstructions. However, we do see that the off-diagonal terms of the covariance matrix are smaller. This reflects the intended effect of this smoothing derivative penalty to stabilise the oscillations between different bin constraints. If we were to use a much finer binning than adopted in Sec.~\ref{sec:results}, then a derivative penalty would need to be used to prevent oscillatory, non-physical reconstructions.

\end{document}